\documentclass{aa}

\usepackage{graphicx,color}
\usepackage{txfonts}
\usepackage{amsmath}
\usepackage{multirow}
\usepackage{natbib,twoopt}
\usepackage[hyphenbreaks]{breakurl}
\usepackage[breaklinks]{hyperref}
\usepackage{placeins}

\DeclareMathAlphabet\mathzapf       {T1}{pzc} {mb} {it}
\bibpunct{(}{)}{;}{a}{}{,}
\definecolor{cobalt}{rgb}{0.06, 0.2, 0.65}
\hypersetup{
  colorlinks,
  citecolor=cobalt,
  linkcolor=cobalt,
  urlcolor=cobalt,
}

\makeatletter
  \newcommandtwoopt{\citeads}[3][][]{\href{http://adsabs.harvard.edu/abs/#3}%
    {\def\hyper@linkstart##1##2{}%
     \let\hyper@linkend\@empty\citealp[#1][#2]{#3}}}
  \newcommandtwoopt{\citepads}[3][][]{\href{http://adsabs.harvard.edu/abs/#3}%
    {\def\hyper@linkstart##1##2{}%
     \let\hyper@linkend\@empty\citep[#1][#2]{#3}}}
  \newcommandtwoopt{\citetads}[3][][]{\href{http://adsabs.harvard.edu/abs/#3}%
    {\def\hyper@linkstart##1##2{}%
     \let\hyper@linkend\@empty\citet[#1][#2]{#3}}}
  \newcommandtwoopt{\citeyearads}[3][][]%
    {\href{http://adsabs.harvard.edu/abs/#3}
    {\def\hyper@linkstart##1##2{}%
     \let\hyper@linkend\@empty\citeyear[#1][#2]{#3}}}
\makeatother

\newcommand{\XMM}{{\em XMM-Newton\,}} 
\newcommand{\UCAM}{{ULTRACAM\,}}

\newcommand{\lta}{\la}

\begin{document}


\title{X-ray and optical observations of the millisecond pulsar binary PSR\,J1431$-$4715}
  \titlerunning{X-ray and optical observations of PSR\,J1431$-$4715}
	\authorrunning{D. de Martino et al.}

\author{D. de Martino\inst{1}
	\and
	A. Phosrisom\inst{2}
	\and
	V. S. Dhillon\inst{3,4}
	\and
	D. F. Torres\inst{5,6,7}
	\and
	F. Coti Zelati\inst{5,6,8}
	\and
	R.P. Breton\inst{2}
	\and
	T.R. Marsh\inst{9}\thanks{Deceased}
	\and
	A. Miraval Zanon\inst{10,11}
	\and
	N. Rea\inst{5,6}
	\and
	A. Papitto\inst{11}
	}

\institute{INAF-Osservatorio Astronomico di Capodimonte, Salita Moiariello 16, I-80131 Naples, Italy\\
         \email{domitilla.demartino@inaf.it}
           \and
	Jodrell Bank Centre for Astrophysics, Department of Physics and Astronomy, The University of Manchester, Manchester M13 9PL, UK
        \and
	Department of Physics and Astronomy, University of Sheffield, Sheffield, S3 7RH, UK
	\and
	Instituto de Astrof{\'{\i}}sica de Canarias, E-38205 La Laguna, Tenerife, Spain
        \and
	Institute of Space Sciences (ICE, CSIC), Campus UAB, Carrer de Can Magrans s/n, E-08193 Barcelona, Spain
        \and
	Institut d'Estudis  Espacials de Catalunya (IEEC), Carrer de Gran Capit\'a 2-4, E-08034 Barcelona, Spain
        \and
	Instituci\'o Catalana de Recerca i Estudis Avan\c cats (ICREA), E-08010 Barcelona, Spain
        \and
	INAF-Osservatorio Astronomico di Brera, Via Bianchi 46, I-23807 Merate (LC), Italy
        \and
	Department of Physics, University of Warwick, Coventry CV4 7AL, UK
        \and
	ASI - Agenzia Spaziale Italiana, Via del Politecnico snc, 00133 Rome, Italy
        \and
	INAF-Osservatorio Astronomico di Roma, Via Frascati 33, I-00040 Monte Porzio Catone (RM), Italy
	}

\date{Received 3 June 2024 / Accepted 2 September 2024}

\abstract{
We present the first X-ray observation of the energetic millisecond pulsar binary
PSR\,J1431$-$4715, performed with \XMM and complemented with fast optical 
multi-band photometry acquired with the \UCAM instrument at ESO-NTT. 
It is found as a faint X-ray source without a significant orbital modulation. This
contrasts with the majority of systems that instead display substantial X-ray orbital
variability.
The X-ray spectrum is dominated by non-thermal emission and, due to the 
lack of orbital modulation, does not favour an origin in an intrabinary shock between the pulsar and
companion star wind. While thermal emission from the neutron star polar 
cap cannot be excluded in the soft X-rays, the dominance of synchrotron 
emission favours an origin in the  pulsar magnetosphere that we describe 
at both X-ray and gamma-ray energies with a synchro-curvature model. 
 
The optical multi-colour light curve folded at the 10.8\,h orbital period is 
double-humped, dominated by ellipsoidal effects, but also affected by 
irradiation. The \UCAM light
curves are fit with several models encompassing direct heating and a cold 
spot, or heat redistribution after irradiation either through convection or 
convection plus diffusion. 
Despite the inability to constrain the best irradiation models, the fits 
provide consistent system parameters, giving an orbital inclination of 
59$\pm6^{\circ}$ and a distance of 3.1$\pm$0.3\,kpc.
The companion is found to be an F-type star, underfilling its Roche lobe 
($f_{\rm RL}=73\pm4\%$), 
with a mass of 0.20$\pm0.04\,\rm M_{\odot}$, confirming the redback status,
 although hotter than the majority of redbacks. 
The stellar dayside and 
nightside temperatures of 7500\,K and 7400\,K, respectively, indicate a 
weak irradiation effect on the companion, likely due to its high 
intrinsic luminosity. Although  the pulsar mass cannot be precisely derived, 
a heavy  (1.8-2.2\,\rm $M_{\odot}$) neutron star is favoured.
}


\maketitle

\section{Introduction}
\label{sec:intro}

Millisecond pulsar (MSP) binaries are compact systems consisting of a
fast spinning neutron star (NS) and a low-mass companion star. Those in a tight
orbit ($\rm P_{orb}\lta$ 1\,d)  are dubbed "spiders" 
\citep{Roberts13} and, depending on the companion mass ($\rm M_c$),  
 are subdivided into "black widows" ($\rm M_c \lta 0.1\,M_{\odot}$)
and "redbacks" ($\rm M_c \sim 0.1-0.4\,\,M_{\odot}$). These old NSs in binaries
are believed to be spun-up to very short spin periods during a previous
Gyr-long phase of mass accretion from an evolved companion. 
According to the recycling
scenario \citep{Alpar82,Backer82} during the accretion phase, MSP binaries were
Low-Mass X-ray Binaries (LMXBs) and turned into radio and gamma-ray pulsars when mass
accretion ceased. The first observational evidence of transitions from a LMXB 
to a rotation-powered pulsar state was found in the MSP PSR\,J1023+0038 \citep{Archibald09}.
However, the subsequent detections of state transitions from and/or to a disc state
in IGR\,J1825$-$2452, in the M28 globular  cluster \citep{Papitto13,linaresetal14}, in 
the Galactic field X-ray source XSS\,J12270$-$4859 
\citep[][]{Bassa14,deMartino14} and again in PSR\,J1023+0038
\citep{Stappers13,Patruno14} imply that
 transitions between accretion-powered and rotation-powered states can
occur on timescales much shorter than secular evolution, likely due to
changes in the mass transfer rate from the companion star and possibly 
controlled by the interplay between the pulsar spin-down power and the 
companion star wind or by its magnetic activity 
\citep[see review][]{PapittodeMartino22}.  

In spiders, the interaction of the relativistic pulsar wind with the 
companion star wind produces an intrabinary shock (IBS), where particles 
are accelerated, evaporating the late-type star \citep{Arons_Tavani93}. 
Depending on the energy of the particles, the IBS cools via inverse Compton or 
synchrotron radiation. Also, 
depending on the pulsar spin down power and companion wind momentum, the IBS
assumes different orientations, wrapping around the companion star in black widows 
or the pulsar in redbacks \citep{Hui15,Wadiasingh18,Kandel19,Sim24}. 
X-ray orbital modulations are indeed observed 
in a number of black widows and redbacks  with different orbital phasing indicating the 
different IBS geometry \citep{Roberts18}. Furthermore redbacks are found to be  more luminous
on average in the X-rays than black widows and isolated MSPs, indicating that a larger fraction of
pulsar wind is intercepted at the shock compared to black widows 
\citep{Lee18,K_Linares23}.

\noindent The late-type companions are generally found to be strongly 
irradiated in black widows,
whilst in redbacks the effect is not always detected
\citep{Breton13,Romani15,Strader19}. The irradiation pattern in both black widows
and  redbacks is not always found to be consistent with direct irradiation 
by the pulsar, sometimes requiring illumination from the IBS as well \citep{Romani16}, or 
even additional heating due to magnetic activity of the companion star \citep{Sanchez17}.

The number of spiders and candidates of both types has largely increased 
recently, thanks to several and deep radio pulsar searches 
 \citep[e.g.][]{Keith10,Han21,MeerKATGPS23,MeerKAT24} and to the gamma-ray {\em Fermi}-LAT 
survey \citep{Abdollahi22}. The latter, in particular,  has allowed 
efficient detection of MSP binaries, which are unaffected at high energies by
eclipse effects from the intrabinary material \citep{Smith23}.
Redbacks are of particular interest because they include the recently discovered
subclass of transitional millisecond pulsar binaries. Among  the 
newly identified redbacks a number of them still lack multi-band
observations to derive the physical parameters of the IBS and the 
companion star.  

In this paper we  focus on the poorly studied MSP binary
PSR\,J1431$-$4715 (henceforth J1431). It was discovered in the  
High Time Resolution Universe Survey (HTRU) by \citet{Bates15} as one of the
fastest (2.01\,ms) and most energetic pulsars with a spin-down power 
$\rm \dot E = 6.8\times 10^{34}\,erg\,s^{-1}$. Its pulsed radio 
emission was found to be  affected by strong  eclipses at the binary
10.8h orbital period. The radio orbital solution classified  
J1431 as a MSP binary with a non-degenerate donor, with a minimum mass of
$\rm M_c = 0.12\,M_{\odot}$. J1431 has also recently been detected at high 
energies by {\em Fermi}-LAT and catalogued as 4FGL\,J1431.4-4711 in the 
4FGL-DR3 12-yr and 4FGL-DR4 14-yr LAT catalogs 
\citep{Abdollahi22,Ballet23}, respectively,
with $\rm L_{\gamma} =1.26-8.11\times 10^{33}\,erg\,s^{-1}$
for a distance range of 1.5-3.8kpc, encompassing distance values 
derived from dispersion measure (DM) \citep{Bates15,Jennings18} and 
{\it Gaia} \citep{Antoniadis21,gaiadr3}\footnote{The {\it Gaia} DR3 geometric distance 
ranges beteen 1.61-2.58\,kpc}. 
A deep search for pulsations has also identified J1431 as a gamma-ray pulsar at the $\sim 7\sigma$ level 
\citep{Bruel19} and thus it is included  in the recent $\rm 3^{rd}$ {\em Fermi}-LAT  catalog 
of gamma-ray pulsars  \citep{Smith23}, joining  the sample of 14 confirmed redbacks 
with radio and gamma-ray detections.  Medium time resolution optical photometry and 
spectroscopy was carried out by \citet{Strader19} that revealed  
the orbital motion of the companion star and found it to be affected by mild 
irradiation. The binary parameters were not constrained due to the degeneracy
between inclination and filling factor. 

\noindent We present here a multi-band analysis of J1431 consisting of an X-ray 
observation of J1431 performed with \XMM, that has allowed the detection 
for the first time of the X-ray emission, as well
as optical multicolour fast photometry acquired at ESO La Silla 
with the \UCAM instrument mounted at the NTT telescope. The
observations are presented in Sect.~\ref{sec:obs}, the X-ray timing and
spectral analysis are reported in Sect.~\ref{sec:xray} and the optical
photometric study in Sect.~\ref{sec:opt}. We discuss the results
in Sect.~\ref{sec:discussion} and compare them with other redback properties.

\section{Observations and data reduction}
\label{sec:obs}

\subsection{The \XMM observation}
\label{sec:xmmobs}

J1431 was observed by \XMM  on 2021 Jan. 21 (OBSID: 0860430101) with the
EPIC-pn camera \citep{Struder01} in the Large Window imaging mode and with the
thin optical blocking filter for an  exposure time of 92.3\,ks and with the EPIC-MOS1 
and MOS2 cameras \citep{Turner01} in Partial Window imaging mode with the thin filter 
for an  exposure time of 93.3 and 94.6\,ks, respectively. The presence
of background flaring activity at the beginning and end of the pointing  
reduced the effective exposures to 78.9\,ks and 84.4\,ks for the EPIC-pn and MOS 
cameras, respectively\footnote{The filtering of the event files was performed creating good time intervals (GTIs)
by setting a cut in the background rate time series without source contributions}.
The Optical Monitor (OM) \citep{mason01} was set in 
fast window mode using the V (5100-5800$\AA$) and UVW1 
(2450-3200$\AA$) filters. Nine OM science windows of $\sim$4400\,s each were
acquired sequentially in the V and UVW1 filters, totalling an exposure 
of 39.6ks in each of them, respectively. 
The log of the observations is reported in Tab.~\ref{tab:tabobs}

\noindent The data were processed and analysed using the \XMM Science Analysis Software
({\sc SAS} v.20.0) with the latest calibration files. The photon arrival
times from EPIC cameras were corrected to the Solar System barycentre 
using the JPL DE405 ephemeris and the nominal position of J1431 reported 
by \citet{Bates15}\footnote{The radio position is fully consistent with the
{\it Gaia} DR3 position \citep{gaiadr3}}. For the EPIC cameras, we
extracted events using a 16" radius circular region centred on the source
and using a background region of the same size located on the same CCD chip. 
To improve the S/N the data were filtered by selecting pattern pixel events up to 
double with zero-quality flag for the EPIC-pn data and up to quadruple for 
the EPIC-MOS data. Background subtracted light curves were produced using 
the {\sc epiclccorr} task in the whole 0.2-12\,keV range and in two
bands covering the 0.2-4\,keV and 4-12\,keV ranges. Given the faintness of the
source, the light curves were binned into  1200\,s and 2400\,s-long intervals, 
respectively. 
The EPIC spectra were extracted in the 0.3-10\,keV range, filtering the events using good
time intervals obtained by selecting low background epochs with the task {\sc gtigen}. 
Response matrix and ancillary files were generated using the tasks {\sc rmfgen} and 
{\sc arfgen}, respectively. The spectra were rebinned with a minimum
of 30 counts in each bin using the {\sc specgroup} routine before fitting. 

The OM V and UVW1 fast window data were processed and background light
curves were generated with the task {\sc omfchain} with a bin time of 400\,s and
1100\,s, respectively. Given the small size of the fast window mode, the centering
of the target was inspected for each window. Due to a drift of the telescope, 
 the target was found to be partially outside the window of the last V band 
observation 
and the first UVW1 band window and  therefore these were disregarded. 
The light curves
were then corrected to the Solar System barycentre in the same manner as the 
EPIC data. The OM V and UVW1 count rates were also converted into AB 
magnitudes using the count rate magnitude conversion available at the \XMM 
SOC\footnote{https:/www.cosmos.esa.int/web/xmm-newton/sas-watchout-uvflux}.

\begin{table*}
\centering
\footnotesize
\flushleft
\caption{The observing log of J1431}
\label{tab:tabobs}
\begin{tabular}{lccccc} 
\hline
Telescope & Instrument & Date & UT(start) & $\rm T_{expo}$\tablefootmark{a} & Average rate (cts\,s$^{-1}$)\tablefootmark{b}\\

\hline

\XMM & EPIC-pn  & 2021-01-21 & 15:59:46 & 84.0\,ks & 8.46$\pm$0.59$\times 10^{-3}$\\
     & EPIC-MOS1 & 	     & 15:28:05 & 84.4\,ks & 2.42$\pm$0.30$\times 10^{-3}$\\
     & EPIC-MOS2 & 	     & 15:28:28 & 84.4\,ks & 2.09$\pm$0.29$\times 10^{-3}$\\
     & OM-V      &           & 15:36:27 & 35.2\,ks & 1.01$\pm$0.02 (17.92$\pm$0.02)\\
     & OM-UVW1  & 2021-01-22 & 04:07:21 & 32.0\,ks & 0.26$\pm$0.01 (20.16$\pm$0.03)\\
\hline
{\em NTT} & \UCAM $u_{\rm s}$,$g_{\rm s}$,$r_{\rm s}$ & 2019-04-12 & 00:45:50 & 1540x20.2\tablefootmark{c}\,s & $u_{\rm s}$=19.18$\pm$0.03, $g_{\rm s}$=17.85$\pm$0.01, $r_{\rm s}$=17.77$\pm$0.01\\ 
          & \UCAM $u_{\rm s}$,$g_{\rm s}$,$r_{\rm s}$ & 2019-04-13 & 03:59:31 & 1024x20.2\tablefootmark{c}\,s & $u_{\rm s}$=19.17$\pm$0.03, $g_{\rm s}$=17.85$\pm$0.01, $r_{\rm s}$=17.78$\pm$0.01\\
          & \UCAM $u_{\rm s}$,$g_{\rm s}$,$i_{\rm s}$ & 2019-04-14 & 02:17:30 & 1203x20.2\tablefootmark{c}\,s & $u_{\rm s}$=19.17$\pm$0.04, $g_{\rm s}$=17.85$\pm$0.01, $i_{\rm s}$=17.67$\pm$0.01\\


\hline
\end{tabular}

\smallskip
\tablefoot{
\tablefoottext{a}{Effective exposure after removal of high background (EPIC) and bad OM centering
in the fast window (OM) or poor data close to twilight (\UCAM). For \UCAM the number of frames and exposure of each frame are reported.}   
\tablefoottext{b}{Average rates in the 0.2-12\,keV range for EPIC cameras. For the OM average rates and AB magnitudes in parenthesis are reported. 
For \UCAM nightly average AB magnitudes in the observed filters are reported.} 
\tablefoottext{c}{One third of frames, each three times the reported exposure (see text), were acquired in the $u_{\rm s}$ filter.} 
}
\end{table*}

\subsection{The \UCAM observations}
\label{sec:ucamobs}

J1431 was observed with the ESO 3.5m New Technology Telescope ({\em NTT}) 
at La Silla (Chile) equipped with the \UCAM photometer \citep{Dhillon07} that allows simultaneous 
high speed three-colour photometry for three consecutive nights from 2019 April
12 to April 14 in relatively good seeing (1-1.5$"$) conditions.
 
\noindent During the first two nights the Super SDSS 
$u_{\rm s}$,$g_{\rm s}$,$r_{\rm s}$ filters
were used, while on the third night the photometry was acquired with the 
$u_{\rm s}$,$g_{\rm s}$,$i_{\rm s}$ filters. The Super SDSS filters are described in \citet{Dhillon21}. 
Single exposure times were 60.6\,s for the $u_{\rm s}$ 
filter and 20.2\,s for the $g_{\rm s}$,$r_{\rm s}$, $i_{\rm s}$ filters. The instrument was set in 
full-frame and no-clear mode giving a dead time between each frame of 0.024\,s
with a GPS time stamp of each frame to an absolute accuracy of 1\,ms. Due
to a rotator flip at transit, the science exposures were stopped for a few minutes. 
The log of  the \UCAM observations is reported in Tab.~\ref{tab:tabobs}.

The data were reduced using the improved reduction pipeline developed for 
{\em HIPERCAM} \citep{Dhillon21}. The images were debiased using master 
bias and dark frames and flat-fielded using a master twilight flat field.
Aperture photometry with variable sized object apertures scaled to the 
seeing in each frame was applied to extract counts. The sky was 
determined from a clipped mean in an annulus around the object aperture 
and subtracted. 
Several comparison stars in the same field of view were also extracted to 
perform differential photometry. Among the comparison stars two
of them have similar colour to J1431\footnote{The comparison stars colours 
were checked against {\it Gaia} DR3}, which then have been used to calibrate 
the target using their magnitudes from the SkyMappper Southern 
Survey DR2 catalogue \citep{Skymapper19}. These have also been checked
against variability and we chose the more stable reference star 
SMSS\,J143149.5-471538.1  with AB magnitudes u=17.61$\pm$0.03, 
g=15.89$\pm$0.01, r=15.50$\pm$0.01 and i=15.261$\pm$0.008.
The light curves in the different bands were inspected against degradation
due to twilight at the end of the observing runs and the affected points
were removed from the light curves. Barycentric correction was
also applied to the time series in the different filters.

\section{Data analysis and results}
\label{sec:analysis}

\subsection{The X-ray emission}
\label {sec:xray}

\begin{figure}
	\includegraphics[width=\columnwidth]{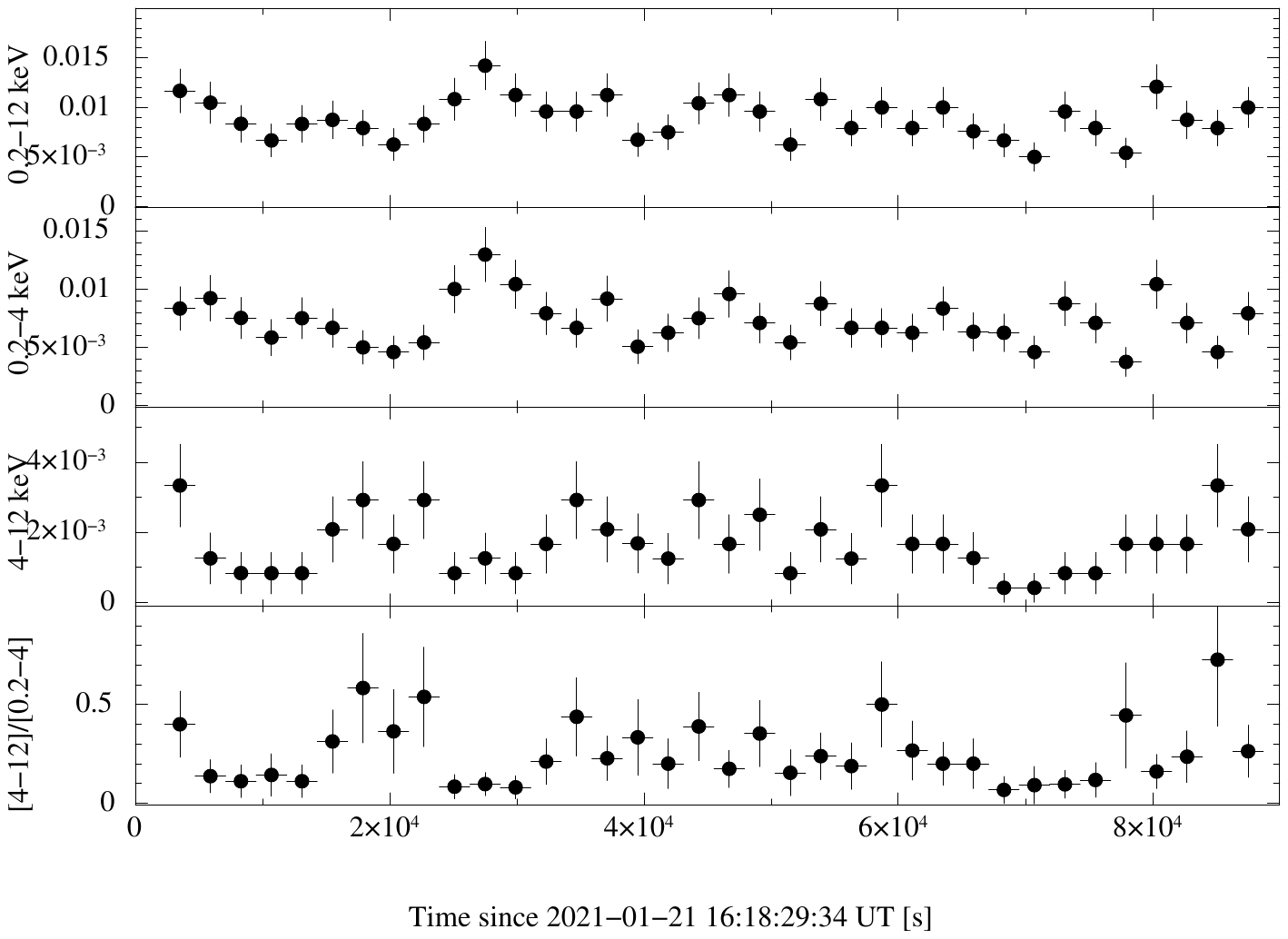}
    \caption{{\it From top to bottom:} The EPIC-pn light curve of J1431 in the 0.2-12\,keV,
0.2-4\,keV, 4-12\,keV ranges and the hardness ratio between the hard and soft bands, 
displayed with a bin size of 2400\,s.}
    \label{fig:pnlc}
\end{figure}

The \XMM observation finds J1431 as a faint source (Tab.~\ref{tab:tabobs}) and
thus we restrict to the more sensitive EPIC-pn data for the timing analysis, 
after inspection of no improvement in the S/N  when summing the rates from 
the three  EPIC cameras.   
In Fig.~\ref{fig:pnlc} the EPIC-pn light curve in the total 0.2-12\,keV is displayed 
together with those in   
the soft 0.2-4\,keV and hard 4-12\,keV ranges. Although the 
length of the observation covers twice the 10.8\,h binary period, there is no
clear orbital modulation in the total band. The soft 0.2-4\,keV  bands may hint
at a possible variability with a marginal increase in the flux at about 28\,ks
from the start of the observation, not observed in the hard (4-12\,keV) range. 
The hardness ratio HR, defined as the ratio of  count rates between the 4-12\,keV and 0.2-4\,keV bands,  
may indicate a softening at that time, but also at earlier and later times as depicted in Fig.~\ref{fig:pnlc}, 
although at these epochs the hard band count rates are consistent with zero given the 
uncertainties on the background subtracted rate.

\begin{figure*}
	\includegraphics[width=\columnwidth]{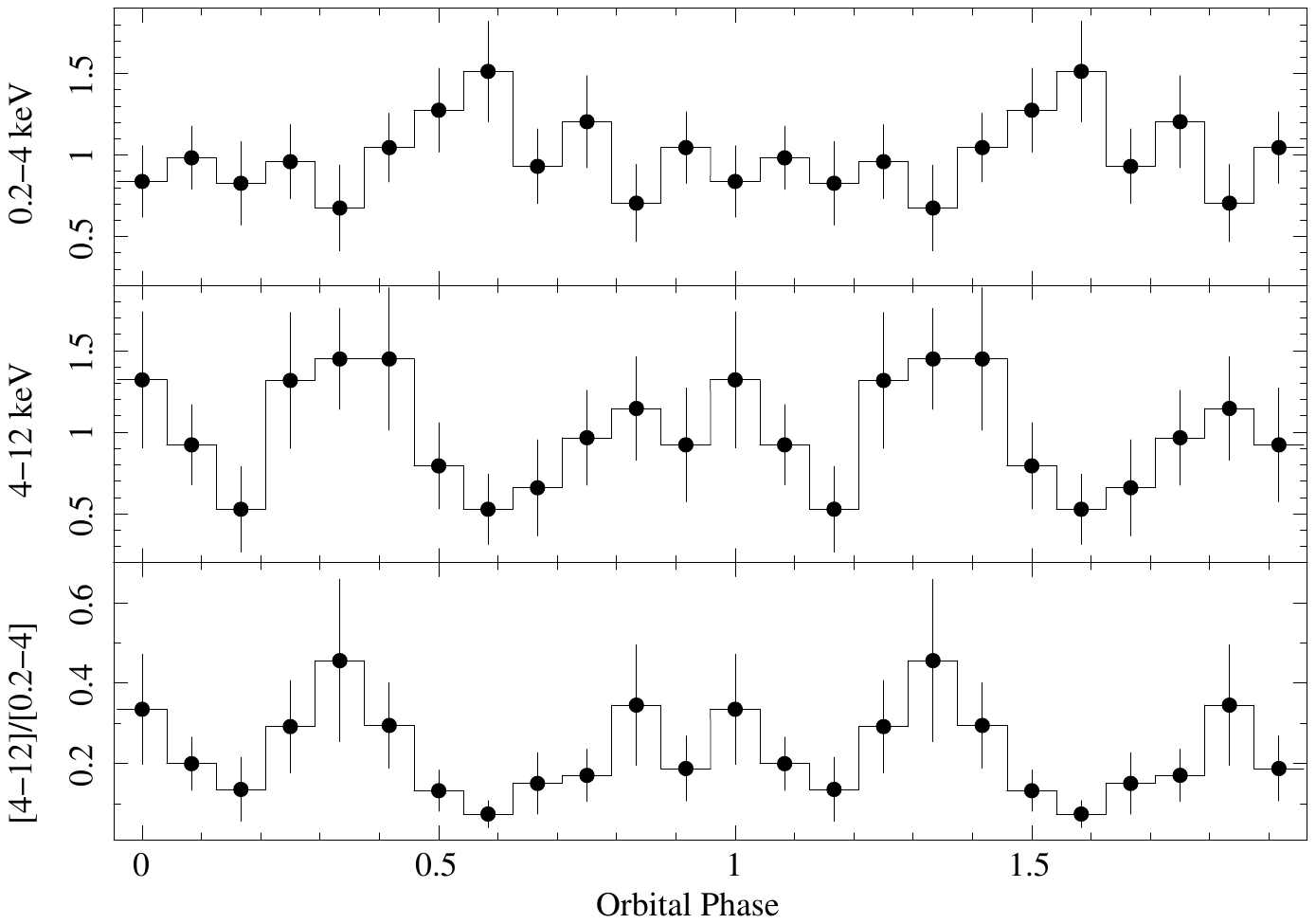}
	\includegraphics[width=\columnwidth]{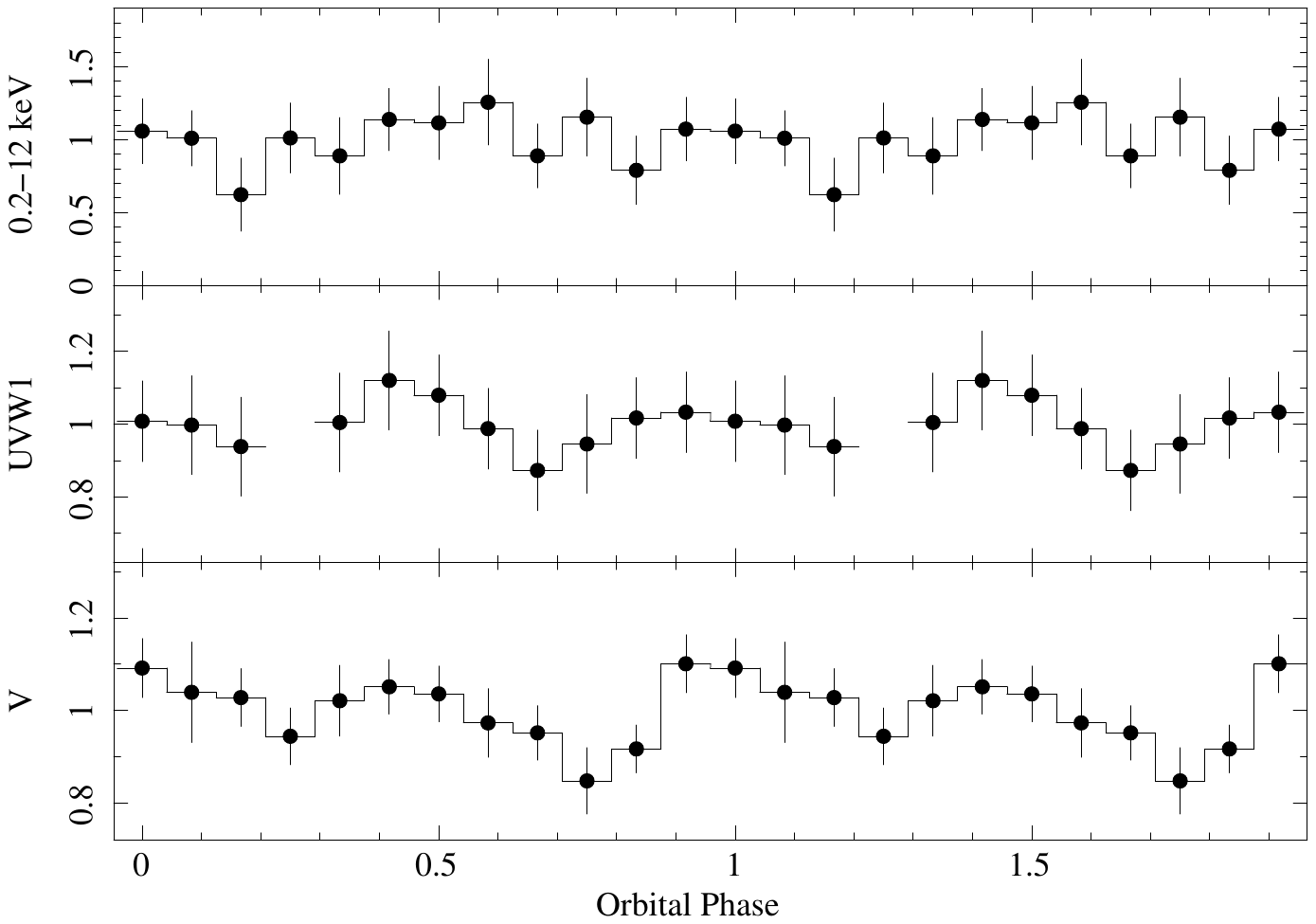}
    \caption{{\it Left:} The EPIC-pn light curves of J1431 in 
the 0.2-4\,keV (upper panel), 4-12\,keV (middle panel) ranges normalised to their average count rates 
and the hardness ratio between the two (lower
panel) folded at the 10.8\,h orbital period evaluated over 12 phase bins. Phase zero corresponds to 
the time of the
ascending node of the pulsar. {\it Right:} The folded light curves in the total 0.2-12\,keV, UVW1 and V 
bands normalised to their average count rates.}
    \label{fig:foldpnom}
\end{figure*}

\noindent The EPIC-pn energy resolved light curves were also folded at the 10.8\,h binary orbital 
period using  the \citet{Bates15} ephemeris (see left panels of Fig.~\ref{fig:foldpnom}), where phase zero corresponds
to the time of the NS ascending node\footnote{The radio solution is highly 
accurate ($d\phi/\phi = 1.15\times 10^{-10}$) allowing the 
folding of the \XMM data}. The soft band  hints to a  variability with
a semi-amplitude of $14\pm12\%$  ($\rm \chi^2_{\nu}/d.o.f. = 0.94/9$), while a fit with a constant 
gives $\rm \chi^2_{\nu}/d.o.f. =0.99/11$. The signficance of an orbital variability estimated with an F-test is only 
at 1.6$\sigma$ confidence level, thus not statistically 
significant. The light curve in the harder (4-12\,keV) range appears to be double-humped
with unequal maxima at $\rm \phi_{orb}\sim$ 0.25-0.4 and $\sim$0.8-1.0. These phases
are close but slightly later than the superior and inferior conjunction of the pulsar, respectively. 
Fitting the hard band orbital light curve with the fundamental and harmonic frequencies, the amplitude is $34\pm 14\%$ 
($\rm \chi^2_{\nu}/d.o.f. = 0.60/9$), while a fit with a constant gives $\rm \chi^2_{\nu}/d.o.f. =1.20/11$. The significance
of the modulation estimated with an F-test is at 4$\sigma$ confidence level. 
The hardness ratios indicate a hardening at these two maxima (bottom left panel in Fig.~\ref{fig:foldpnom}). 

\noindent  On the other hand, the light curve in the 
total range does not reveal significant orbital modulation 
(right upper panel in Fig.~\ref{fig:foldpnom}) being consistent with a constant ($\rm \chi^2_{\nu}/d.o.f. = 0.50/11$)
likely due to the higher count rate in the soft than in the hard band. 
While a detailed comparison with the optical emission is performed in
 Sect.\,\ref{sec:opt}, the simultaneous UV and optical photometry  
(Fig.~\ref{fig:foldpnom}, right bottom panels) 
shows a weak double-humped variability with amplitudes of $8\pm6\%$ in the 
UVW1 band and $8\pm3\%$ in 
the V band with unequal maxima at the ascending and descending nodes of the NS,  although an F-test gives 
significance at 3$\sigma$ and 2.5$\sigma$ levels.
The different behaviour will be discussed in Sect.\,\ref{sec:discussion}.

\begin{figure}
	\includegraphics[width=\columnwidth]{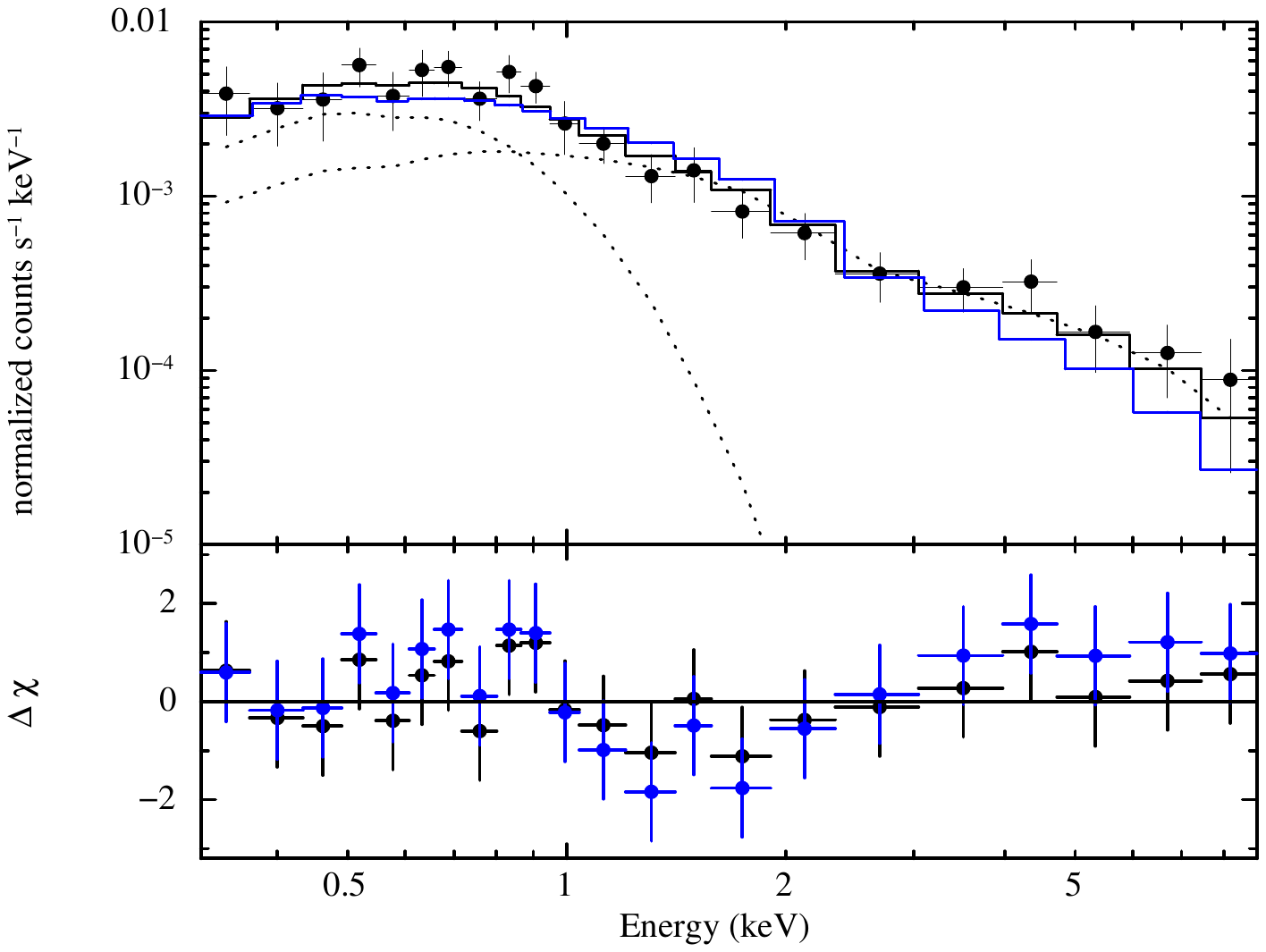}
\caption{{\it Upper panel:} The average 0.3-10\,keV spectrum of J1431 (black points). 
Only the EPIC-pn is shown for clarity, fitted with an absorbed power law (solid blue line) and a 
composite model (solid black line) consisting of  an absorbed power law and a thermal blackbody 
(dotted black lines) with parameters reported in the text. {\it Lower panel:} The residuals of the two
fits are shown  with blue and black points, respectively.}
    \label{fig:pnspec}
\end{figure}

\noindent The X-ray  0.3-10\,keV spectrum averaged over the whole observation is 
featureless with no sign of a cut-off. 
The EPIC-pn and MOS1,2 spectra were analysed using {\sc xspec} 
package \citep{arnaud96} and fitted together. We first  adopt a simple model consisting of an 
absorbed power law, {\sc const*tbabs*powerlaw}, where {\sc tbabs} accounts
for the ISM absorption with abundances from \cite{wilms00}, and {\sc const}
accounts for intercalibration among the three instruments. This model  gives a reasonable fit 
($\rm \chi^2_{\nu}/d.o.f. = 1.01/26$), with a power law index 
$\alpha =1.63_{-0.18}^{+0.29}$ (uncertainties are 90$\%$ confidence level
on one interesting parameter) and an unconstrained neutral hydrogen 
column density $\rm N_H \leq 8\times10^{20}\,cm^{-2}$.
%
This upper limit is roughly consistent with    
the total ISM hydrogen column density in the direction of the source
($\rm N_{H,ISM} = 1.1\times 10^{21}\,cm^{-2}$ \citep{HI4PI16})  and the column
density 
from the dispersion measure DM=59.35$\rm pc\,cm^{-3}$ of \citet{Bates15}, 
$\rm N_{H,DM} = 1.8(6)\times 10^{21}\,cm^{-2}$,
using the empirical relation of \citet{he13}, as well as with the column 
density 
%
%
%
of $\rm \sim 1.1\times 10^{21}\,cm^{-2}$,
derived from the recent $\rm 3D-N_{H}$ tool of \citet{Doroshenko24} that combines dust maps and distance to estimate
the optical reddening and X-ray absorption adopting the same distance range.   
Fixing the column density to 
$\rm N_{H,ISM} = 8\times 10^{20}\,cm^{-2}$ the fit provides similar results 
($\rm \chi^2_{\nu}/d.o.f. = 1.08/27$) for the power law index 
$\alpha= 1.88\pm0.19$ (see Fig.\,\ref{fig:pnspec})\footnote{The results do not change adopting a column 
density of $\rm 1.0\times 10^{21}\,cm^{-2}$}. 
The unabsorbed flux in the 0.3-10\,keV
range is $\rm 2.6\pm0.3\times 10^{-14}\,erg\,cm^{-2}\,s^{-1}$.
%
%
At the distances of 1.53-3.8\,kpc the corresponding luminosity is 
$\rm L_X = 0.7-5.0\times 10^{31}\,erg\,s^{-1}$. 

Although the simple power law model is formally acceptable, we also included a 
blackbody component, using {\sc bbodyrad } to inspect
whether the presence of the thermal emission of the NS is required. We do not 
attempt to use a NS atmosphere model (e.g. {\sc nsa}) given the low statistics
of the spectra.  
Keeping fixed the hydrogen column density as before we find a power law index
$\alpha=1.31_{-0.38}^{+0.36}$, a blackbody temperature 
$\rm kT_{BB}= 0.15\pm0.04$\,keV and normalisation $\rm N_{BB}= 1.12_{-0.72}^{+2.63}$
($\rm \chi^2_{\nu}/d.o.f. = 0.83/25$) (see Fig.\,\ref{fig:pnspec}). 
The latter implies an emitting radius 
$\rm R_{BB}\sim$100-740\,m for a distance in the range of 1.53-3.8\,kpc.
%
%
The 0.3-10\,keV unabsorbed 
fluxes of the two components result in: $\rm F_{BB}=6.4_{-0.5}^{+0.3}\times 
10^{-15}\,erg\,cm^{-2}\,s^{-1}$ and $\rm F_{pow}=2.6_{-0.2}^{+0.1}\times 
10^{-14}\,erg\,cm^{-2}\,s^{-1}$. The total luminosity in the 0.3-10\,keV range
then results  $\rm L_X \sim 0.9-5.55\times 10^{31}\,erg\,s^{-1}$ for the
above distance range.
The inclusion of the blackbody using  
an F-test is however significant only at a confidence level of 2.5$\sigma$.

J1431 displays a similar spectrum to  other MSP binaries in
the rotation-powered state, with a power
law index within the range $\alpha$=1.0-1.9. 
The X-ray emission is therefore dominated by synchrotron radiation with
a luminosity consistent with that derived in redbacks and higher than
those of black widows \citep[see][]{Lee18,yap23,K_Linares23}. 
Although not required from the fits, a blackbody component with 
temperature of $\sim$0.15\,keV and size of a few hundreds of meters is reasonably
consistent with those found in MSPs, where the NS thermal
emission from the heated pole is detected 
\citep{zavlin06,Bogdanov06,Bogdanov11}.  
The ratio of the unabsorbed fluxes between the putative thermal component and 
the power law in the 0.3-10\,keV is $\sim$0.25 
and the bolometric luminosity of the thermal component 
in the range $\rm 2.3-15.0\times 10^{30}\,erg\,s^{-1}$ are also consistent
with those found in MSPs \citep{zavlin06,Bogdanov11}. 
Whether the non-thermal emission originates in the IBS or from the pulsar magnetosphere 
will be discussed in  Sec.\,\ref{sec:discussion}. 

\subsection{The optical light curves}
\label{sec:opt}

\begin{figure}
	\includegraphics[width=2.6in,angle=-90]{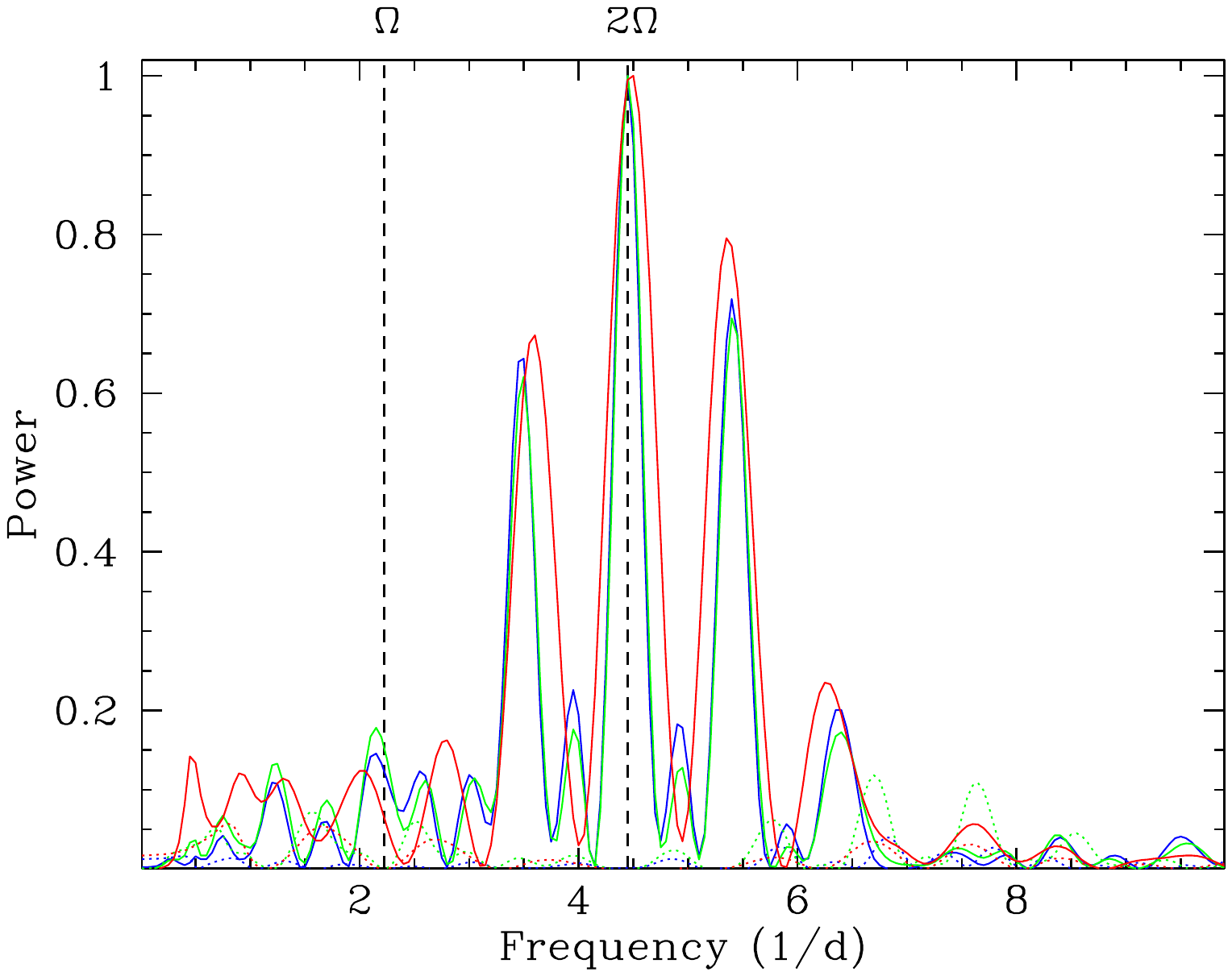}
    \caption{Lomb-Scargle periodogram of the \UCAM light curves (solid lines) 
in $u_{\rm s}$ (blue), $g_{\rm s}$ (green) and $r_{\rm s}$ (red) 
filters displaying the strongest
peak at twice the orbital frequency (dashed vertical line). The periodograms
of the residuals of the sinusoidal fits  in each colour
are reported with a dotted line. }
    \label{fig:powspec}
\end{figure}

The \UCAM time series clearly show a low-amplitude periodic variability
in the $u_{\rm s}$,$g_{\rm s}$,$r_{\rm s}$,$i_{\rm s}$ bands. Using the longer time coverages in the 
$u_{\rm s}$,$g_{\rm s}$,$r_{\rm s}$  filters, the Lomb-Scargle periodograms \citep{lomb76,scargle82} 
reveal the presence of strong  power at the harmonic of the 10.8\,h orbital period 
(see Fig.\,\ref{fig:powspec}). 
A sinusoidal fit composed of the fundamental and harmonic  
to the light curves gives a period 
$\rm P_{u_s}^{\Omega}$=0.45024(22)\,d, $\rm P_{g_s}^{\Omega}$=0.449499(48)\,d and
$\rm P_{r_s}^{\Omega}$=0.44950(12)\,d fully consistent with the more precise orbital
period of 0.4497391377(7)\,d derived from radio timing by \citet{Bates15}. 
The modulation amplitudes in the four bands are  $\rm A_{u_s}^{\Omega}$ = 0.0193(8)\,mag, 
$\rm A_{u_s}^{2\Omega}$ = 0.0734(2)\,mag; $\rm A_{g_s}^{\Omega}$= 0.0215(1)\,mag, 
$\rm A_{g_s}^{2\Omega}$ = 0.0682(1)\,mag, $\rm A_{r_s}^{\Omega}$=0.0153(2)\,mag, 
$\rm A_{r_s}^{2\Omega}$ = 0.0630(1)\,mag and 
$\rm A_{i_s}^{\Omega}$= 0.0117(3)\,mag, $\rm A_{i_s}^{2\Omega}$ = 0.0601(2)\,mag, 
where for the $i_{\rm s}$ band the orbital frequency of \citet{Bates15} has been adopted.

\begin{figure}
	\includegraphics[width=\columnwidth]{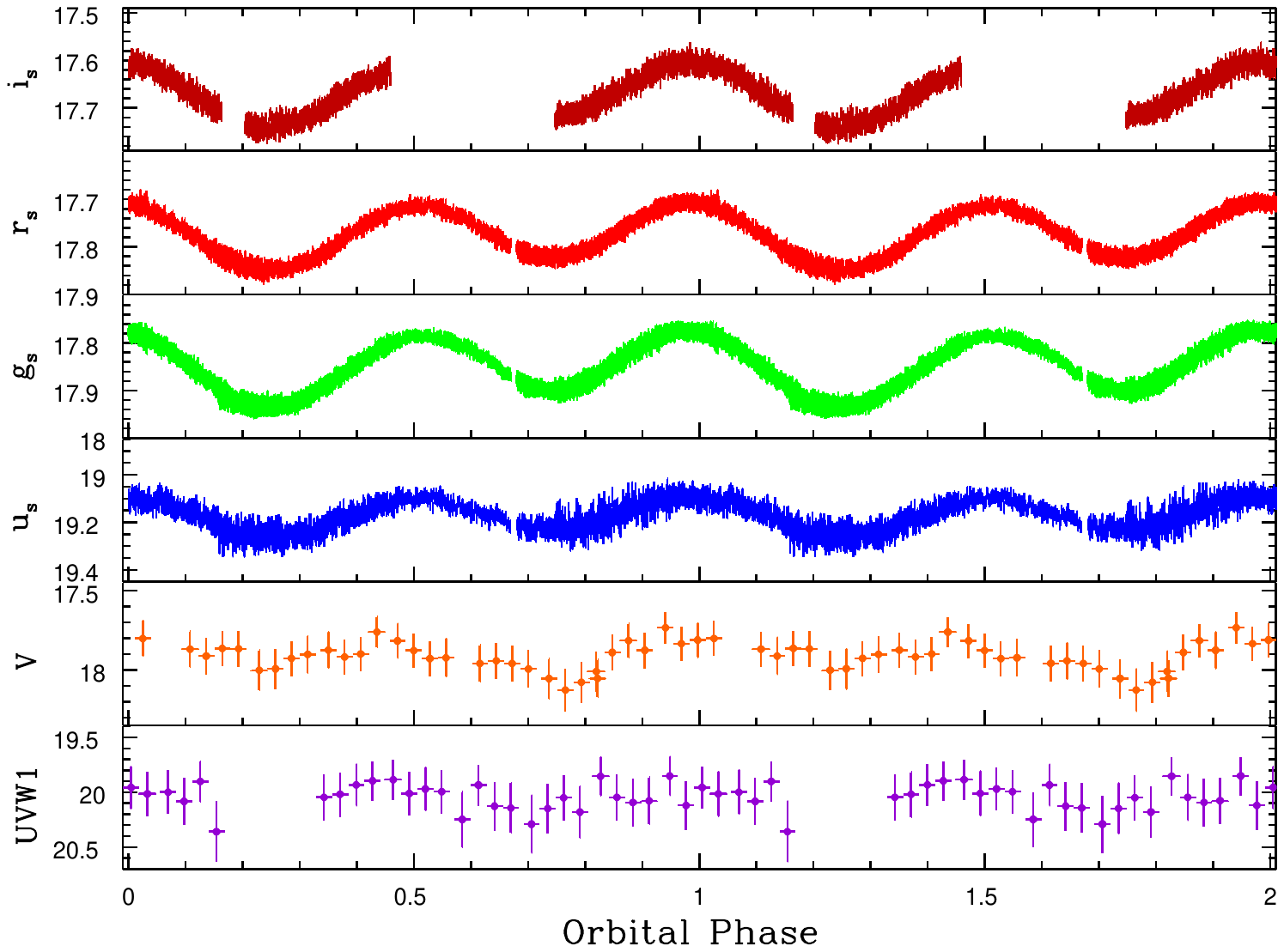}
    \caption{{\it From bottom to top:} The observed orbital light curves in the OM UVW1 and V filters acquired in 2021 
and in the \UCAM  $u_{\rm s}$, $g_{\rm s}$, $r_{\rm s}$ and $i_{\rm s}$  filters acquired in 2019. Ordinates are AB magnitudes.} 
    \label{fig:foldopt}
\end{figure}

The observed light curves, including the OM UVW1 and V band filters 
folded at the orbital period using the accurate ephemeris of 
\citet{Bates15} are shown in Fig.\,\ref{fig:foldopt}. 
The double-humped shape clearly demonstrates the presence of ellipsoidal
variations due to the tidal distortion of the companion star, which however
are not symmetrical and do not show equal minima.  These minima occur
at $\rm \phi_{orb} \sim$ 0.25 and 0.75. Also slight phase shifts are observed
at the two maxima, one anticipating at $\phi_{orb}=1.0$ and the other lagging at $\phi_{orb}=0.5$.
The wavelength dependent amplitudes also hint at colour effects, 
likely due to irradiation as pointed out by \citet{Strader19}. 
It is worth of note the different shape of the V band orbital modulation observed in 2021 with
respect to the \UCAM light curves acquired in 2019, possibly indicating
a change in irradiation efficiency, although the lack of a deeper minimum
at $\rm \phi_{orb} \sim$ 0.25 could also be due to the lower  accuracy of 
the OM photometry.  We then limit ourselves to the analysis of
the high time resolution \UCAM data.

\subsubsection{Modeling of the optical light curves}
\label{sec:icarus}

The \UCAM light curves, complemented with the
radial velocity curve obtained from the H$_{\alpha}$ absorption line 
by \citet{Strader19} for the companion star,
were analysed with the {\sc icarus}\footnote{https://github.com/bretonr/Icarus} 
binary modeling code \citep{breton12} 
to constrain system parameters and the temperature profile of the companion. 
This software is used along with {\sc ATLAS9} stellar atmosphere grids \citep{Castelli_Kurucz03}
to construct a photometric grid of synthetic atmosphere models,
which in turn are folded through the \UCAM filter transmissions to 
obtain the observed flux at a given distance. The {\sc MultiNest} 
nested sampling algorithm  \citep{Feroz19} was used to explore the 
parameter space, which provides the Bayesian evidence (Z) of a model.

\begin{table*}
\footnotesize
\centering
\renewcommand{\arraystretch}{0.9}
\caption{The posterior parameters obtained from the fits of our multi-colour
light curves and radial velocity curve from \citet{Strader19}.}  
\label{tab:tabicarus}
\begin{tabular}{l|c|c|c|c|c|c|c|c}
\hline
Model & DH & C & D+C & DH+Spot & DH & C & D+C & DH+Spot \\

\hline

$\beta$ & 0.25 & 0.25 & 0.25 & 0.25 & 0.08 & 0.08 & 0.08 & 0.08 \\

\hline
&  &   &  &  &  & & &   \\

\multicolumn{9}{c}{Fitted Parameters} \\

E(B-V) & $0.45_{-0.03}^{+0.03}$ & $0.45_{-0.03}^{+0.03}$ & $0.45_{-0.03}^{+0.03}$ & $0.45_{-0.03}^{+0.02}$ & $0.18_{-0.03}^{+0.03}$ & $0.19_{-0.03}^{+0.03}$ & $0.19_{-0.03}^{+0.03}$ & $0.17_{-0.03}^{+0.03}$ \\
$K_{\rm c}$ (km s$^{-1}$) & $278_{-3}^{+2}$ & $279_{-3}^{+2}$ & $279_{-3}^{+2}$ & $280_{-3}^{+1}$ & $278_{-2}^{+2}$ & $279_{-2}^{+1}$ & $279_{-2}^{+1}$ & $278_{-2}^{+2}$ \\
$d$ (kpc) & $3.9_{-0.2}^{+0.2}$ & $3.9_{-0.2}^{+0.2}$ & $3.9_{-0.2}^{+0.2}$ & $4.1_{-0.2}^{+0.2}$ & $2.9_{-0.3}^{+0.2}$ & $3.3_{-0.2}^{+0.2}$ & $3.2_{-0.2}^{+0.2}$ & $3.0_{-0.2}^{+0.2}$ \\
$i$ ($^\circ$) & $48_{-1}^{+1}$ & $48_{-1}^{+1}$ & $48_{-1}^{+1}$ & $45.5_{-0.9}^{+0.9}$ & $62_{-5}^{+7}$ & $55_{-3}^{+3}$ & $56_{-3}^{+3}$ & $60_{-3}^{+5}$ \\
T$_{\rm base}$ (K) & $10200_{-70}^{+70}$ & $10210_{-60}^{+70}$ & $10210_{-70}^{+60}$ & $10200_{-70}^{+70}$ & $7540_{-70}^{+60}$ & $7600_{-100}^{+100}$ & $7670_{-100}^{+90}$ & $7510_{-90}^{+60}$ \\
T$_{\rm irr}$ (K) & $6700_{-100}^{+100}$ & $6700_{-100}^{+100}$ & $6700_{-100}^{+100}$ & $6700_{-100}^{+100}$ & $3860_{-70}^{+70}$ & $4000_{-70}^{+70}$ & $4040_{-70}^{+60}$ & $3880_{-60}^{+60}$ \\
$f_{\rm RL}$ & $0.77_{-0.01}^{+0.02}$ & $0.77_{-0.01}^{+0.02}$ & $0.77_{-0.01}^{+0.02}$ & $0.79_{-0.01}^{+0.01}$ & $0.70_{-0.03}^{+0.03}$ & $0.74_{-0.02}^{+0.03}$ & $0.73_{-0.02}^{+0.02}$ & $0.71_{-0.02}^{+0.02}$ \\
C$_{\rm ampl}$ (W K$^{-1}$ m$^{-2}$) & - & $-35000_{-2000}^{+1000}$ & $-35000_{-2000}^{+1000}$ & - & - & $-15000_{-800}^{+800}$ & $-15000_{-1000}^{+1000}$ & - \\
$\kappa_{\rm diff}$ (W K$^{-1}$ m$^{-2}$) & - & - & $300_{-300}^{+800}$ & - & - & - & $2000_{-1000}^{+2000}$ & - \\
$\phi_{\rm spot}$ ($^\circ$) & - & - & - & $67_{-3}^{+2}$ & - & - & - & $-71_{-2}^{+2}$ \\
$\theta_{\rm spot}$ ($^\circ$) & - & - & - & $132_{-9}^{+10}$ & - & - & - & $111_{-6}^{+9}$ \\
R$_{\rm spot}$ ($^\circ$) & - & - & - & $16_{-2}^{+2}$ & - & - & - & $9_{-2}^{+3}$ \\
T$_{\rm spot}$ (K) & - & - & - & $2800_{-700}^{+2000}$ & - & - & - & $-500_{-400}^{+200}$ \\
&  &   &  &  &  & & &   \\

\multicolumn{9}{c}{Derived Parameters} \\
$q$ & $10.44_{-0.1}^{+0.08}$ & $10.47_{-0.1}^{+0.06}$ & $10.47_{-0.1}^{+0.06}$ & $10.50_{-0.1}^{+0.05}$ & $10.43_{-0.08}^{+0.07}$ & $10.48_{-0.08}^{+0.05}$ & $10.48_{-0.08}^{+0.05}$ & $10.44_{-0.08}^{+0.07}$ \\
M$_{\rm p}$ (M$_\odot$) & $2.9_{-0.2}^{+0.2}$ & $3.0_{-0.2}^{+0.2}$ & $3.0_{-0.2}^{+0.2}$ & $3.4_{-0.2}^{+0.2}$ & $1.8_{-0.3}^{+0.3}$ & $2.2_{-0.2}^{+0.3}$ & $2.2_{-0.2}^{+0.3}$ & $1.8_{-0.2}^{+0.2}$ \\
M$_{\rm c}$ (M$_\odot$) & $0.28_{-0.02}^{+0.02}$ & $0.29_{-0.02}^{+0.02}$ & $0.28_{-0.01}^{+0.02}$ & $0.32_{-0.02}^{+0.02}$ & $0.17_{-0.03}^{+0.03}$ & $0.21_{-0.02}^{+0.03}$ & $0.21_{-0.02}^{+0.02}$ & $0.18_{-0.02}^{+0.02}$ \\
T$_{\rm day}$ (K) & $9620_{-70}^{+70}$ & $9620_{-60}^{+60}$ & $9620_{-60}^{+60}$ & $9730_{-90}^{+100}$ & $7450_{-70}^{+60}$ & $7500_{-100}^{+100}$ & $7570_{-100}^{+90}$ & $7420_{-90}^{+60}$ \\
T$_{\rm night}$ (K) & $9410_{-70}^{+60}$ & $9420_{-60}^{+60}$ & $9420_{-60}^{+60}$ & $9400_{-70}^{+70}$ & $7370_{-70}^{+60}$ & $7400_{-100}^{+100}$ & $7490_{-100}^{+90}$ & $7350_{-90}^{+60}$ \\
R$_c$ (R$_\odot$) & $0.68_{-0.02}^{+0.02}$ & $0.69_{-0.02}^{+0.02}$ & $0.68_{-0.02}^{+0.02}$ & $0.72_{-0.02}^{+0.02}$ & $0.54_{-0.05}^{+0.04}$ & $0.61_{-0.03}^{+0.04}$ & $0.59_{-0.03}^{+0.03}$ & $0.56_{-0.04}^{+0.03}$ \\
&  &   &  &  &  & & &   \\
\multicolumn{9}{c}{Goodness of fit} \\
log Z           & -2097.5 & -1606.0 & -1611.6 & -1568.2 & -1891.3 & -1451.5 & -1453.7 & -1419.0 \\
N$_{dof}$       & 8703 & 8702 & 8701 & 8699 & 8703 & 8702 & 8701 & 8699 \\
$\chi^{2}_{\rm \nu}$ & 0.591 & 0.529 & 0.528 & 0.524 & 0.566 & 0.511 & 0.511 & 0.507 \\
\hline
\end{tabular}
\tablefoot{The different
models with $\beta$=0.25 and $\beta$=0.08 are grouped from left to right:
direct heating (DH), convection (C), diffusion+convection (D+C) and direct 
heating with a spot (Spot), applying  gravity darkening after 
irradiation (post-IGD). The derived
parameters are listed below the fit parameters. These are the mass ratio $q$,
the pulsar M$_{\rm psr}$ and companion M$_{\rm c}$ mass, the day T$_{\rm day}$
and night T$_{\rm night}$ temperatures, which are obtained averaging
the temperature over the surface elements of the day and night side hemispheres (i.e.
in the plane of the orbit) according to 
$\rm T^4 = \sum_i (T_i^4 \, s_{i,{\rm eff}})$/$\rm \sum_i s_{i,{\rm eff}}$, with $\rm T_i$
the temperature and $\rm s_{i,{\rm eff}}$ the projected area of a surface element. 
$\log{\rm Z}$ is the Bayesian evidence of the fit with more positive values indicating better
fits. The uncertainties are the 1$\sigma$ confidence intervals.}
\end{table*}

The precise pulsar radio-timing solution derived by \citet{Bates15} was adopted 
to fix the input parameters: the time of the ascending node of the pulsar 
($\rm T_{asc}$= BMJD 55756.1047771), the orbital period 
($\rm P_{orb}$=0.4497391377\,d) and the projected 
semimajor axis of the pulsar orbit ($x=a_{\rm psr}\,sin i/c$=0.550061\,lt-s). 
The pulsar mass function $f(\rm M_{psr})$ relates to the system parameters
as $f(\rm M_{psr})=\rm M_{psr}$\,$sin^{3}i/(1+1/q)^{2} = 
q^{3}\,x^{3}\,4\,\pi^{2}/G\,\rm P_{orb}^{2}$ where $q$=$\rm M_{psr}/M_{c}$ is the  
pulsar-to-companion mass ratio, $i$ is the binary inclination. Since the mass 
function is related to the companion star radial velocity amplitude $K_{c}$ 
as $f(\rm M_{psr})$=$\rm P_{orb}\,K_{c}^{3}\,/2\,\pi\,G$, the parameters to 
be fit  are $i$, $\rm K_{c}$.
The {\sc icarus} code constructs a stellar surface whose radius is 
parametrized by the Roche lobe filling factor $f_{\rm RL}$, 
defined as the ratio between the
radius of the companion in the direction of the pulsar and the distance 
between the companion centre of mass and the inner Lagrangian point 
($\rm L_{1}$). $f_{\rm RL}$ is then another fit parameter.
The  surface stellar temperature $\rm T_{base}$ is defined as the 
temperature of the 
pole prior to irradiation and,  accounting for gravity darkening, 
$\rm T_{base}$ is
multiplied by $(g/g_{\rm pole})^{\beta}$, where $g$ is the acceleration 
gravity and
$\beta$ is the gravity darkening  exponent.  This parameter is fixed at 
$\beta=0.08$ and $\beta=0.25$, which are appropriate for convective and 
radiative envelopes, respectively \citep{lucy1967}. 
The effect of the pulsar heating is represented by the 
irradiating temperature $\rm T_{irr}$, defined such that a flux 
$\rm \sigma\,T_{irr}^{4}$ at the star centre of mass is received 
at a distance $a$ from the pulsar. With the assumption that the irradiating
flux is immediately thermalised and re-radiated \citep{breton12}, each surface 
element at a distance $r$ from the pulsar, with its normal vector 
forming an angle $\theta$ with respect the vector pointing to the pulsar, 
has a temperature such that 
$\rm T^{4} = [T_{base}^4$\,$(g/g_{\rm pole})^{4\beta}$ $\rm 
+ T_{irr}^{4}$\,$\cos\,\theta\, (a/r)^{2}]^{1/4}$.
This model assumes direct heating (DH) from the pulsar and applies gravity 
darkening prior to irradiation, here termed pre-irradiation gravity darkening (pre-IGD) \citep{Breton13}.
Gravity 
darkening applied after irradiation and heat redistribution on the companion
surface, here termed post-irradiation gravity darkening (post-IGD), 
as done in \citet{Romani16,Romani21} and \citet{Dodge24}, has also been investigated. 
However the presence of 
asymmetries and shifts in the light curves of J1431, already noticed by 
\citet{Strader19}, cannot be accounted for by the simple DH approach. 
These features
can instead be modeled either by adding hot or cold spots or by including 
heat redistribution  within the outer layers of the star by convection 
only (C) and/or diffusion and convection (D+C)  
\citep{kandel_romani20,voisin20}. These models have also been inspected by 
applying both pre-IGD and post-IGD. 

The {\sc icarus} code also encodes as fit parameters the optical 
interstellar reddening defined as $E(B-V)$ in the direction of J1431 
determined in each band using dust maps and extinction 
vectors from \citet{Chiang23} as well as the source distance $d$. 
Hence the fit parameters for all these models are: E(B-V), $d$, $\rm K_{c}$, 
$i$, $f_{\rm RL}$, $\rm T_{base}$, $\rm T_{irr}$. For the models accounting 
for heat 
redistribution with convection (C) and diffusion plus convection (D+C),  
additional fit parameters are the convective strength parameter C$_{\rm ampl}$,  
which determines the profile of the convective wind 
and the diffusion parameter $\kappa_{\rm diff}$, which describes the amplitude of 
a linear diffusion effect \citep[see][]{voisin20}. In the case of spot model 
a single surface hot or cold spot is added to the DH model with fitted 
temperature difference T$_{\rm spot}$, angular size R$_{\rm spot}$, and position angles 
$\theta_{\rm spot}$, $\phi_{\rm spot}$ \citep[see][]{Clark21,Stringer21}.

The main priors for the fit parameters were choosen such that they are 
physically or geometrically  motivated.
Among them, the distance of J1431 whose prior is constructed combining
the ranges derived by  \citet{Bates15}, \citet{Jennings18}, 
\citet{Antoniadis21} and {\it Gaia} DR3 (see Sec.\,\ref{sec:intro}), but allowed to vary from 0.1 to 20\,kpc. 
Also a Gaussian prior is applied to the interstellar 
extinction E(B-V), centred at 0.166 with standard deviation of 0.16 but allowed
to vary over a larger range from 0 to 1. The base and irradiation temperatures
were constrained to lie between 2100\,K and 12000\,K, and between 2300\,K 
and 15000\,K, respectively.
For the binary inclination angle $i$, a uniform prior in $\cos\,i$ has been 
adopted. As for the Roche lobe filling factor $f_{\rm RL}$, a uniform prior has
been adopted between 0.2 and 1.0. 
The spectroscopic $\rm K_{c}$ constraints are provided by the observed radial
velocity curves obtained from cross-correlation of the $\rm H_{\alpha}$ absorption line in the
optical spectrum of J1431 by \citet{Strader19}. Here the {\sc icarus} code 
simulates spectra at specific orbital phases and the radial velocities of the model are 
compared to the observed values \citep[see also][]{Dodge24}.    

\begin{figure*}
	\includegraphics[width=\linewidth]{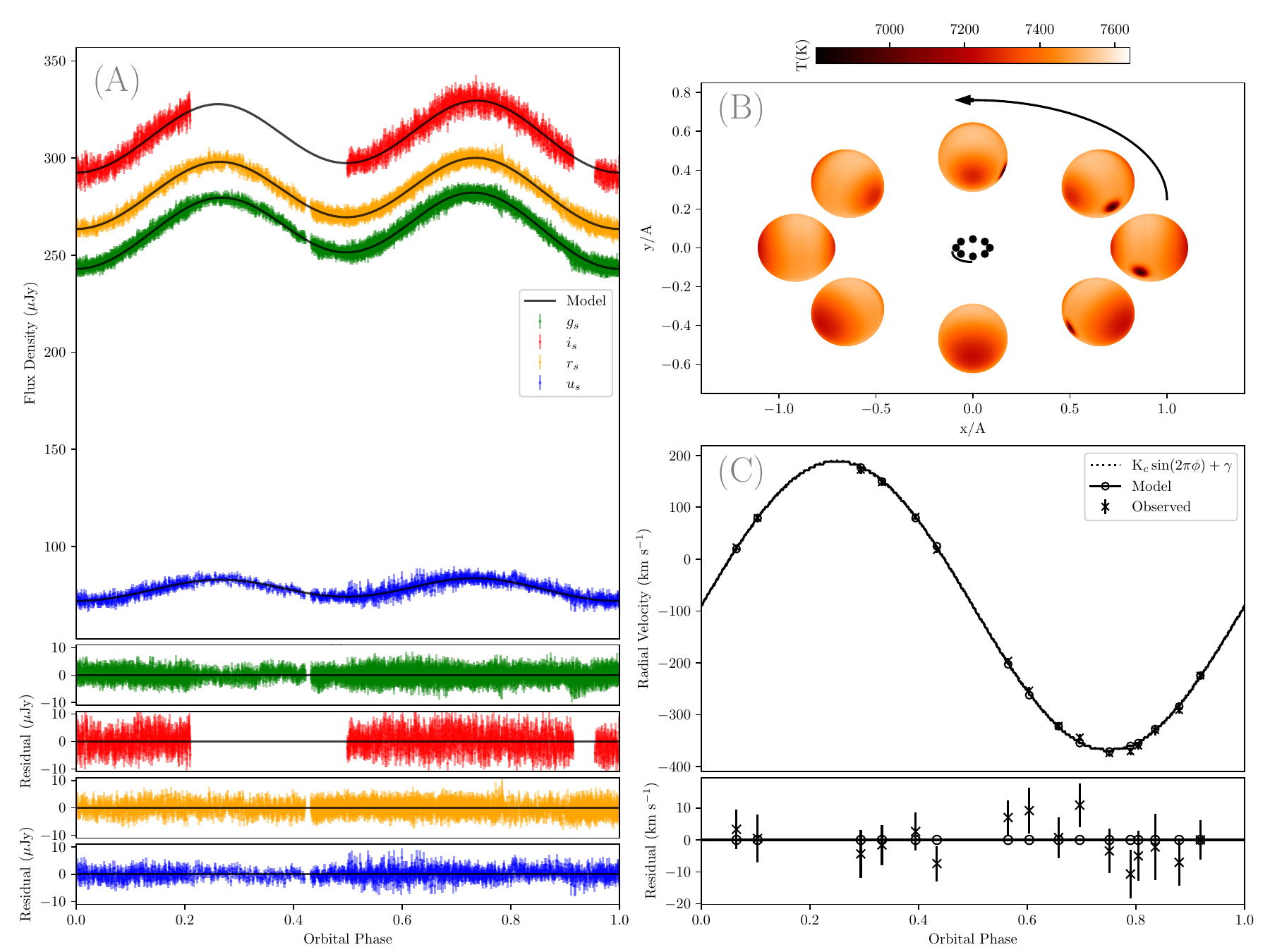}
    \caption{{\it Panel A:} The \UCAM light curves shown with 
different colours fitted with the DH model and a cold spot overlaid with solid 
black lines. The residuals are shown in the lower panels. 
{\it Panel B:} The surface temperature map 
along the orbit. {\it Panel C:} The observed radial velocity curve of 
J1431 obtained by \citet{Strader19} (black dots) and the ones predicted by 
the model (empty dots) connected with a solid line together with a sinusoidal 
fit (dotted line). The residuals are displayed in the lower panel.} 
    \label{fig:spot}
\end{figure*}

Among the several model fits, those with post-IGD prescription outperform the pre-IGD in all cases, 
except identical results are found for the the DH+Spot model adopting either 
pre-IGD or post-IGD heat redistribution prescription.
In Tab.~\ref{tab:tabicarus} the results of the different 
modelling are reported for the post-IGD prescription only and for the two values of the 
gravity darkening exponent $\beta$. Models with $\beta$=0.25 have much lower 
Bayesian evidence (log\,Z) with respect to those adopting $\beta$=0.08, and hence are discarded. 
Here we note that the transition temperature from convective to radiative envelopes 
falls in the range $\sim 7000-7900$\,K  \citep{rafert80,alencar97}, justifying the investigation with both $\beta$ values. 
The fitted parameters with $\beta=0.08$  are found to be all consistent 
within 1$\sigma$ for those models with  
higher Bayesian evidence, namely that with convection (C), with diffusion and convection (D+C) 
and with direct heating (DH) with a cold spot. 
A slight preference for the DH with a cold spot is found. This model is shown in 
Fig.\,\ref{fig:spot} where the fitted 
multi-colour light curves (panel A) are displayed together with the surface 
temperature map of the companion (panel B) as well as the observed and 
model radial velocity curves (panel C). Corner plots for the best  three
models (C), (D+C) and DH+spot with post-IGD are reported
in Fig.\,\ref{fig:cornerc}, Fig.\,\ref{fig:cornerdc} and 
Fig.\,\ref{fig:cornerspot}, respectively. 
The similarities of the results obtained with these three  best fit models provide 
evidence that the derived parameters are reliably constrained despite the
inability of determining whether heat redistribution models or a spotted
surface represent the physical state of the companion star.

The distance of J1431 has not been well constrained 
by several methods \citep{Bates15,Jennings18,Antoniadis21,gaiadr3}, allowing 
for a wide range from 1.5\,kpc to 3.8\,kpc. Here we find consistency for all
the three models, which, adopting the higher and lower bounds of the derived parameters, 
provide  a distance to J1431 of 3.1$\pm$0.3\,kpc, well within the above
range and that we adopt in the forthcoming section. The interstellar extinction 
E(B-V)=0.17-0.19, giving $\rm A_{V}$=0.52-0.59
is also consistent with the upper limit to the 
hydrogen column density found from the X-ray spectral fits (see 
Sec.\,\ref{sec:xray}) using the \citet{guver_ozel09} relation. 
The binary system parameters are discussed in Sec.\,\ref{sec:discussion}.

\section{Discussion}
\label{sec:discussion}

\subsection{A mildly irradiated companion}
\label{sec:companion}

The parameters obtained with the {\sc icarus} code for the three  best fit models
give consistent values within 1$\sigma$ for the binary inclination,  
the companion Roche lobe filling factor and mass. By averaging among the three models we
obtain  $i=59\pm6^\circ$, $f_{\rm RL}=73\pm4\%$, $\rm M_{c}=0.20\pm0.04\,M_{\odot}$, which
are broadly consistent with the preliminary results obtained by 
\citet{Strader19} but more accurate. These values confirm the redback status of J1431. 

\noindent The companion surface is found to have  hemisphere-averaged nightside and 
dayside temperatures  $\rm T_{night}\sim$7400\,K and 
$\rm T_{day}\sim$7500\,K, both higher than previously estimated 
($\sim$6500-6600\,K) by \citet{Strader19}.  We here note that this lower temperature 
estimate is based on spectra covering a narrow spectral range between 5500-6730\,$\AA$
encompassing only H$_{\alpha}$ and not a wider wavelength range as our multi-band photometry. 
We also verified whether the extinction prior could affect the temperature determination adopting
a flat prior for the reddening. The posterior distributions of the fit are found similar to those 
adopting a Gaussian prior, indicating that the reddening does not affect the temperature determination.
The spectral type of the  companion would then be an early-F star,
 much hotter than the majority of redbacks, which are instead found in the range of 2800-6300\,K \citep{turchetta23}, with only
PSR\,J1816+4510 found to be extreme (16000\,K) \citep{kaplan13,turchetta23}.
The companion of J1431 could be an outlier due to an unusual binary evolution,  
or hot redback companions are still to be discovered. This shows that the characterization of the optical companions of newly discovered
redbacks is an important aspect of investigation.
The irradiating temperature 
$\rm T_{irr}\sim 3900$\,K  implies an equivalent irradiation 
luminosity $\rm L_{irr}= 4\,\pi\,a^2\,\sigma\,T_{irr}^4 
\sim 7.8\times 10^{33}\,erg\,s^{-1}$, where $a=2.17\times 10^{11}$\,cm 
having adopted $a_{\rm psr}\,\sin\,i$ from \citet{Bates15} and the derived 
values for the inclination $i=59^{\circ}$ and
mass ratio $q$=10.4.  This luminosity when compared
to the spin-down power $\rm \dot E = 6.8\times 10^{34}\,erg\,s^{-1}$ from
\citet{Bates15} or that derived in Sec.\ref{sec:ns} below,  gives
an efficiency $\rm \eta_{irr}=L_{irr}/\dot E  \sim 12-13\%$, 
not unusual in rebacks, although at the low end of 
irradiation efficiencies \citep[see][]{Breton13,yap23}. To estimate 
the heating effect
on the companion we consider the expected heating impinging on the companion 
in the case of an isotropic pulsar wind $\rm L_{heat} = f_{\Omega}\,\dot E$
where $\rm f_{\Omega}$ is the geometric factor $\sim 0.5\,(1-\cos\theta)$ with 
$\theta= \arctan\, (\rm R_{c}$/$a$), neglecting albedo. Adopting the derived companion
radius $\rm R_c =0.6\,R_{\odot}$ and the above value for $a$, 
$\rm f_{\Omega}$=0.009 and then 
$\rm L_{heat} = 6.1\times 10^{32}\,erg\,s^{-1}$. The ratio between the 
expected heating and the companion unheated luminosities,
$\rm L_{c,base} = 4\pi R_{c}^2\,\sigma\,T_{base}^4 = 4\times
10^{33}\,erg\,s^{-1}$, then   results in  
$\rm \eta_{heat}=L_{heat}/L_{c,base}$= 0.15. This indicates that
the companion luminosity is about 6.5 times larger than the heating, 
hence implying weak irradiation  as indeed evidenced by the observed 
double-humped light curves, indicating that they are dominated by ellipsoidal effects.
The fact that MSP binaries with intrinsic companion  
luminosities larger or comparable to that expected from pulsar irradiation
show negligible heating effects was already pointed out by 
\citet{Hui15} and it is further supported by the results 
by \citet{turchetta23} on a larger number of spiders with a wide 
range of orbital periods. 
The long orbit of J1431 and hence large
separation together with the high companion luminosity can explain the
lack of strong irradiation in this system.

%

\subsection{A heavy neutron star}
\label{sec:ns}

 Noteworthy is the derived  pulsar mass $\rm M_{psr}$ ranging from 1.8 to 2.2
$\,M_{\odot}$, considering all three models. Although we are unable to 
discriminate among the various models employed, which limits the precision to which this 
parameter is constrained, a massive NS in J1431 is favoured.
In the recent years, NS masses in spiders have been subject of several investigations 
\citep[see][and references therein]{linares19,Strader19,Clark21}. Comparing the NS mass derived for J1431 with the latest compilations of 
NS masses in \citep{Kennedy22,Rocha23,Dodge24,sen24}, J1431 joins the 
group of massive NSs, which includes the redbacks PSR\,J2215+5135, PSR\,J1622$-$0315 and PSR\,J1816+4510 and the 
black widows PSR\,J1653$-$0158, PSR\,J1810+1744, PSR\,J0952$-$0607 and 
PSR\,JB1957+20.  Noteworthy is the case of PSR\,J1622$-$0315, a mild irradiated redback potentially hosting a massive
NS \citep{sen24}. These systems are particularly attractive because they 
allow lower limits be set on the maximum NS mass and hence have 
important implications for the dense matter equation of state (EoS).

The spin-down power of the pulsar, defined as 
$\rm \dot E=4\,\pi^2\,I\,\dot P/P^3$, where P and $\rm \dot P$ are the spin period
 and its derivative  and
$\rm I = 2/5 \,M_{NS}\,R_{NS}^2$  is the momentum of inertia of the NS,
was estimated by \citet{Bates15} as $\rm \dot E = 6.8\times 10^{34}\,erg\,s^{-1}$, 
adopting the canonical NS moment of inertia $\rm I=10^{45}\,g\,cm^{2}$
and correcting the spin period derivative for the Shklovskii effect.
The proper motion of J1431 is now better constrained by 
{\it Gaia} DR3 \citep{gaiadr3} $\mu=18.7(2)\,\rm mas\,yr^{-1}$, which 
allows a more accurate determination of the Shklovskii effect. 
We then correct the observed spin period derivative $\rm \dot P$ for 
this effect:
$\rm \dot P_{Sh}/P$ = $v_t^2/(c\,d)$, where $v_t$ is the transverse velocity, 
and $d$ the distance. Adopting $d =3.1\pm$0.3\,kpc, we derive
$\rm \dot P_{corr} = \dot P_{obs} - \dot P_{Sh} = 8.7(4) 
\times 10^{-21}\,s\,s^{-1}$, where $\rm P_{obs}=1.411\times10^{-20}\,s\,s^{-1}$ is the observed
spin period derivative by \citet{Bates15}.
Given the evidence of a higher NS mass than 1.4$\rm M_{\odot}$ (see Tab.~\ref{tab:tabicarus}) , 
we adopt conservatively the lower limit of $\rm M_{psr}=1.8\,M_{\odot}$ and, lacking
knowledge of EoS,  
a NS radius of 10\,km \citep[see also][]{steiner13,han20,riley21}, giving
 $\dot E = \rm 5.8-6.4\times 10^{34}\,erg\,s^{-1}$, slightly lower 
than the previous determination. It is not too far from the value
of $\rm 5.2\pm0.3 \times 10^{34}\,erg\,s^{-1}$  derived
by \citet{K_Linares23} adopting the canonical moment of inertia 
$\rm I=10^{45}\,g\,cm^{2}$ and the lower {\it Gaia} distance of $\sim$1.6-2.6\,kpc. 

Further observations, such as optical spectroscopy, would be desirable
to better investigate the irradiation model and hence constrain the 
NS mass.

\subsection{An X-ray dim redback }
\label{sec:dimx}

 Our first deep X-ray observation of J1431 has detected a faint X-ray source  without signficant variability
at its orbital period. The X-ray emission in the soft (0.2-4\,keV) is unmodulated and in the hard (4-12\,keV) 
range only a hint at 4$\sigma$ level of a double peaked variability is found. This contrasts with the majority
of spiders displaying substantial X-ray orbital variations, which are often found double-peaked 
due to Doppler boosting of the IBS synchrotron emission. In redbacks the non-thermal emission 
is Doppler boosted at inferior conjunction of the NS and de-boosted at superior conjunction. 
The separation of the two peaks and their intensity strongly depends on the bulk Lorentz factor, 
viewing angle and shock radii \citep{Wadiasingh17}. In black widows the maximum of X-rays occurs 
at superior conjunction of the NS, while in the majority of rebacks it is the opposite 
\citep{Roberts18}. This has been interpreted as different geometry of the IBS, which wraps the pulsar/companion in 
redbacks/black widows, respectively  depending to the momentum ratio between the companion and the pulsar 
winds \citep{Romani16,Wadiasingh17}.

In J1431 the companion is found to be luminous, suggesting a higher wind momentum and it is conceivable that
the IBS wraps the pulsar as the majority of redbacks. In this case an orbital modulation is expected and 
affected by Doppler boosting at inferior conjunction of the NS. 
The lack of an orbital modulation in J1431 could arise if the system is viewed at low inclination angles. However
the optical analysis reveals an intermediate binary inclination of $\sim$60$^{\circ}$.
A low inclination is also not favoured by the presence of long radio eclipses at superior conjunction of the pulsar, 
lasting $\sim$ 0.3 in orbital phase \citep{Bates15}. 
However, the possibility that an IB shock dominates the X-ray emission in J1431, not producing an X-ray orbital modulation
cannot be discarded, but determining the properties of the IB shock is not an easy task given the dependence on the binary inclination,
the shock location, opening angle and bulk velocity ($\beta$=v/c) of accelerated particles.
In the thin-shell approximation \citep{canto96}, generally adopted for spiders, for a moderate inclination angle,
shadowing of the companion is not expected to produce signficant obscuration of the shock emission if the shock is located
close to the pulsar and if most of the emission originates close to the shock nose and the opening angle is small.  
Near the shock nose, the bulk velocity is low and Doppler-boosting is not efficient \citep{Wadiasingh17}. 
Given the lack of higher energy coverage above 10\,keV that would allow the  inspection of a 
spectral break to estimate the shock location as well as the lack of coverage of the radio eclipses at low frequencies 
\citep{Bates15} to derive the maximum shock opening angle, it is not possible to verify this hypothesis.

Adopting the derived distance $d=3.1\pm0.3$\,kpc, 
the X-ray luminosity ranges between $\rm 1.4-2.1\times 10^{31}\,erg\,s^{-1}$ in the 
2-10\,keV band, or between $\rm 2.1-3.0\times 10^{31}\,erg\,s^{-1}$ in the 
0.5-10\,keV range. When compared the X-ray luminosity to those observed in spiders, 
J1431 is located between the redbacks and black widows and in 
particular among low X-ray luminosity redbacks \citep{Lee18,yap23,K_Linares23}. 
The efficiency in converting spin-down power to X-ray luminosity in J1431 is 
$\eta_{x}= L_{x}/\dot E \sim 0.02-0.04\%$ (2-10\,keV) or 0.03-0.05$\%$ (0.5-10\,keV).
Although caution should be taken adopting empirical relations, 
this luminosity ratio locates J1431 in between  the so-called "inefficient" and "efficient"  IBS 
tracks derived by \citet{K_Linares23}.
All this indicates a very weak contribution of the IBS, if any. 

\noindent Noteworthy, J1431 not only joins the low X-ray luminosity redbacks but also shares with 
these systems a double-humped optical light curve, dominated by ellipsoidal variability.
Hence, it is very likely that the weak companion heating is not due to 
X-ray irradiation. Indeed our time-resolved X-ray light curve does not
allow us to conclude whether the non-thermal emission originates in an IBS.


The spectral analysis indicates that the emission is dominated by non-thermal radiation, described
with a power-law model with index 1.3-1.9, similar to those observed in spider pulsars 
\citep{Lee18,K_Linares23}. Although formally not required by the spectral fits,
a thermal component at $\rm kT_{BB} \sim$ 0.15\,keV 
originating at the heated NS polar cap ($\rm R_{BB}\sim 180-660$\,m for the
distance range 2.8-3.4\,kpc derived in Sec.\ref{sec:icarus}) 
could be also present. Similar values for this thermal component are found in other spiders 
\citep{zavlin06,Bogdanov06,Bogdanov11},  contributing $\sim20\%$ to the total X-ray flux.
Hence, it is also possible that the dominant X-ray non-thermal emission originates in the NS magnetosphere.
In this case the X-ray emission would be pulsed at the spin period of the NS, but the X-ray data were acquired in 
imaging  mode  with a temporal resolution of 47.7\,ms, preventing any search of spin pulses. Worth of note is that
searches for X-ray spin pulsations in the rotation-powered state of the transitional MSP XSS\,J12270-4859 led negative
detection even using high time resolution data \citep{Papitto15}.

The gamma-ray emission of J1431 is typical of pulsars with a 
spectral shape modelled with a  power law with a superexponential 
cutoff \citep{Ballet23,Smith23}
giving a 100\,MeV-100\,GeV flux of $\rm 4.7\times 10^{-12}\,erg\,cm^{-2}\,s^{-1}$ 
and a  gamma-ray luminosity for a distance range of 2.8-3.4\,kpc  
$\rm L_{\gamma} = 4.4-6.5\times 10^{33}\,erg\,s^{-1}$.
%

The ratio of  gamma-ray luminosity
to spin down power  $\rm \eta_{\gamma} = L_{\gamma}/\dot E \sim 7-11\%$  
indicates an efficiency of conversion consistent with the median of $10\%$ of 
redbacks \citep{Strader19}. 
The gamma-ray emission in pulsar binaries is generally dominated by the pulsar 
magnetospheric radiation
and should not be modulated at the orbital period for moderately low 
binary inclinations. 
However if gamma-rays are also produced in
an IBS an orbital modulation should be observed such as in the bright systems 
XSS\,J12270$-$4853, PSR\,J2339$-$0533 and PSR\,J2039$-$5617 \citep{Clark21,Sim24}, 
in which also the X-rays
are strongly modulated at the orbital period \citep{Roberts18,deMartino15,deMartino20}. 
In J1431  there is no firm evidence of variability in the X-rays and searches in the gamma-rays
do not provide evidence of either an orbital  modulation or eclipses \citep{Clark23}.
The X-ray (0.5-10\,keV) to gamma-ray flux ratio  $\rm F_{x}/F_{\gamma}\sim 0.5\%$ is lower than 
that found in  most redbacks and similar to PSR\,J1908+2105 ($\sim 0.6\%$), surpassing only 
 PSR\,J1622$-$0315 ($\sim 0.3\%$) and PSR\,J1816+4510 ($\sim 
0.05\%$) \citep{Strader19}. It is then 
plausible that the gamma-ray emission in J1431 originates in the pulsar wind.

\begin{figure}
        \includegraphics[width=\columnwidth]{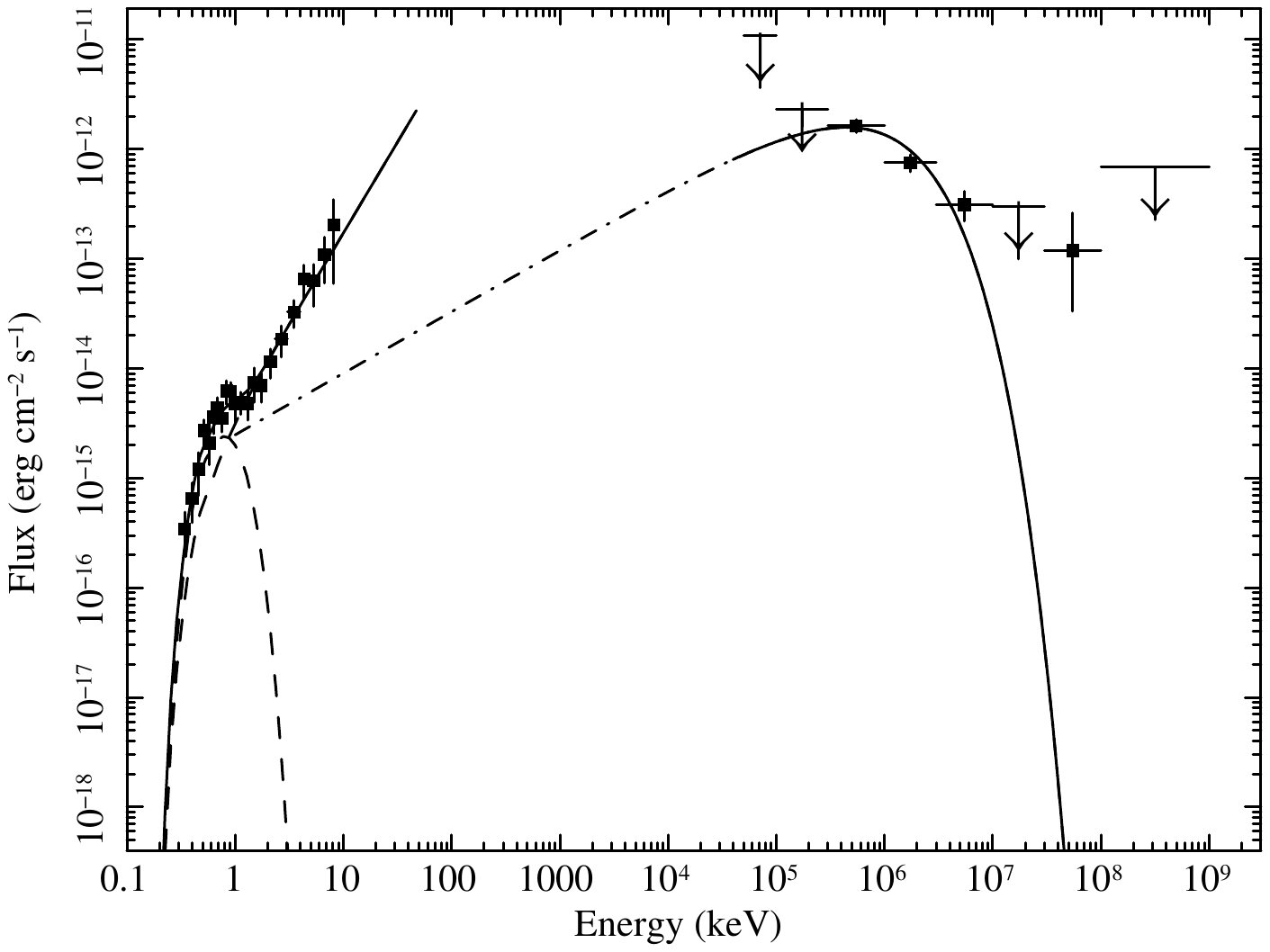}
    \caption{The observed high energy X-ray and gamma-ray SED overlaid with the X-ray composite 
blackbody plus power law model and the superexponential cutoff model from the 3$^{rd}$ {\em Fermi}-LAT pulsar
catalog \citep{Smith23}. Dashed lines represent the single components of the X-ray model and dotted-dashed line
is the extrapolation of the {\it PLEC4} gamma-ray model}
    \label{fig:xraygamma}
\end{figure}

We constructed the broad band X-ray and gamma-ray spectral energy distribution (SED) using our X-ray
spectral analysis adopting the composite blackbody and power law model and the {\em Fermi}-LAT 
SED as retrieved from the recent 3$^{rd}$ pulsar catalog \citep{Smith23} along with the 
typical pulsar superexponential cutoff
power law model  {\it PLEC4} with  peak energy $\rm E_{p} = 0.44\pm0.24$\,GeV, spectral slope $\Gamma$=2.63$\pm$0.16
spectral curvature  $d$=0.8$\pm$0.3 and superexponential index fixed at the
canonical value $b$=2/3 due to the source faintness (see details in \citet{Smith23}). 
The SED (Fig.~\ref{fig:xraygamma}) shows that
the X-ray spectrum, when extrapolated to higher energies, is higher than that predicted by the  
{\it PLEC4} model that, however, cannot be reliably evaluated at lower energies due to the fixed $b$ parameter
at the canonical value. 

\begin{figure*}
        \includegraphics[width=\columnwidth]{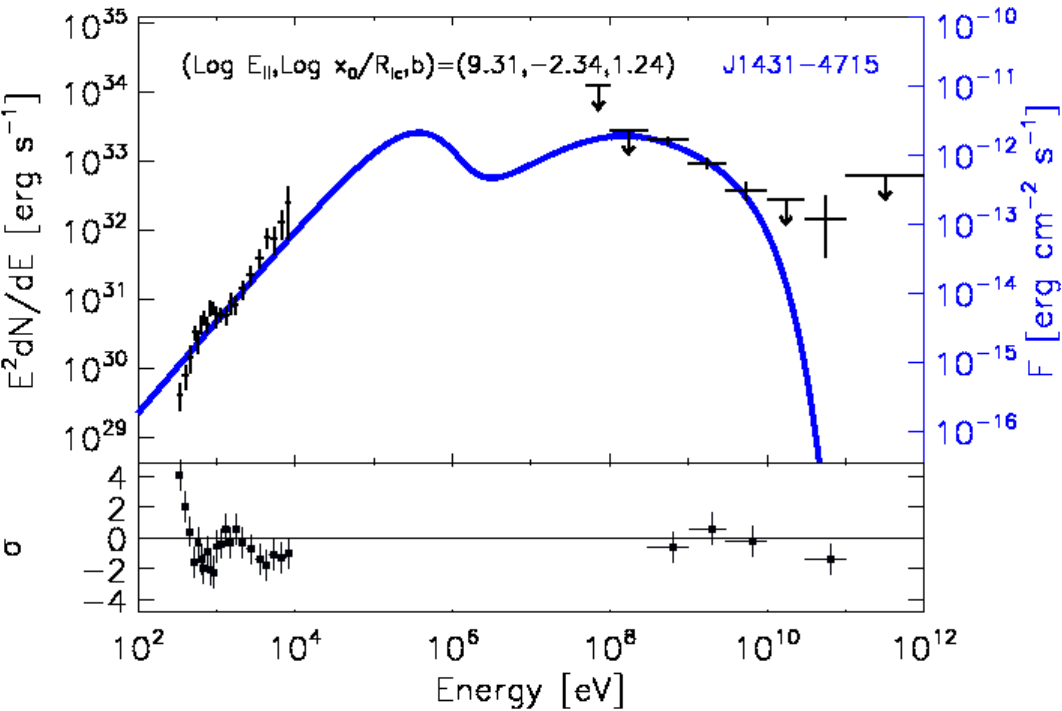}
        \includegraphics[width=\columnwidth]{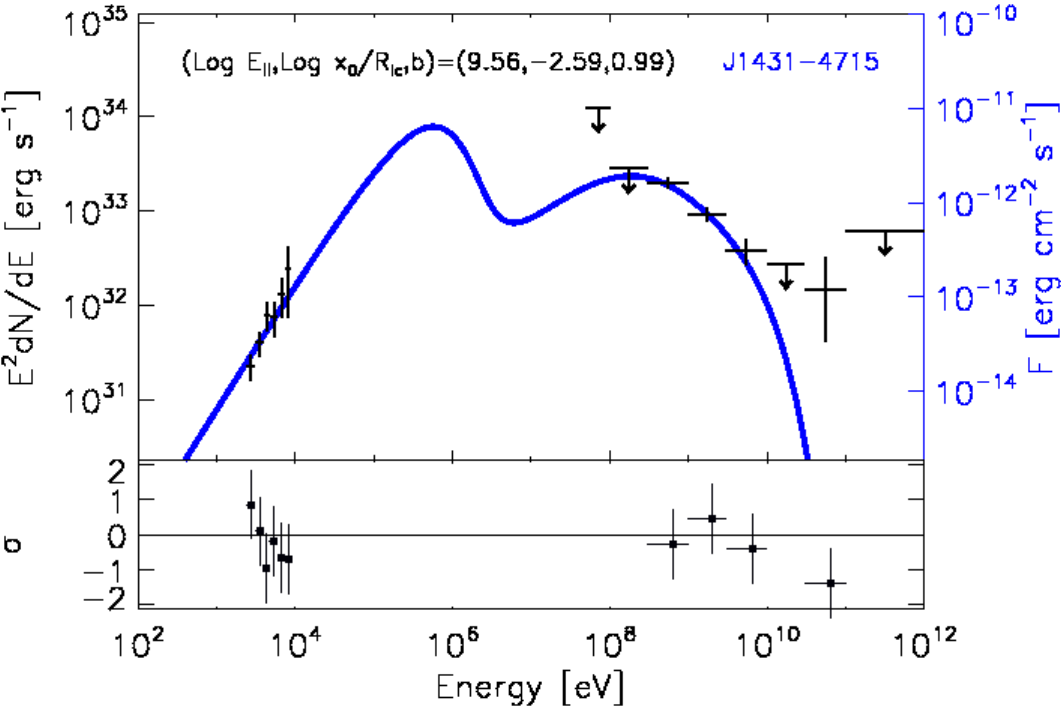}
    \caption{{\it Left:} The X-ray and gamma-ray SED (black points) 
displayed together with the synchro-curvature model fit (blue solid line) 
using the three parameters described in the text. The residuals of the model with respect to
the data are shown in the lower panel. {\it Right:} The same SED but fitted above 2\,keV to avoid 
possible contamination of thermal emission from the NS. In both panels, the ordinates are expressed in 
$\rm erg\,s^{-1}$ for a distance of 3.1\,kpc (left) and in $\rm erg\,cm^{-2}\,s^{-1}$ (right).} 
    \label{fig:sc-fit}
\end{figure*}


We then explored whether a significant part of the gamma-ray and X-ray 
emission could be the direct result of the pulsar magnetospheric emission. 
Curvature and synchrotron emission from accelerated particles in the magnetospheric
gaps or reconnection in the current sheet producing synchrotron radiation at high energies 
are possible mechanisms at the origin of non-thermal pulses. The synchro-curvature emission model 
developed by \cite{Torres18} to describe the high energy spectra of pulsars is based on the assumption
that near the light cylinder of a pulsar with determined spin period and period derivative there is
a gap with a significant component of the electric field parallel to the magnetic lines and this field
accelerates particles. The model follows the particle time evolution, 
solving the equation of motion and balancing acceleration and losses. 
It has been successfully applied to the high energy spectra
of gamma-ray and  X-rays detected pulsars \citep{Torres19} and has been 
further updated 
with improved treatment of particle injection to generate
synthetic spectra and gamma-ray light curves of pulsars 
\citep{Iniguez22,Iniguez24}.
We hence attempt to describe the observed broad-band SED of J1431 with the updated version of the 
synchro-curvature model. This allows us to compare the results with those  obtained for other gamma-ray 
pulsars detected. It also has the advantage that it is a rather austere spectral model
having only  three main physical parameters and a normalisation, which has proven 
useful to represent X-ray and gamma-ray spectra
of pulsars, the appearance of sub-exponential cutoffs at high energies,
or the flattening of the X-ray spectra at soft energies.
The parameters of the model are the electric field parallel to the 
magnetic field lines, $\rm E_{||}$, assumed  constant throughout an accelerating region;
the magnetic gradient $b$,  such that the local magnetic field 
$\rm B= B_s (x/R_s)^{-b}$, where $\rm B_s$
is the surface polar magnetic field estimated from radio timing 
($\sim 2\times 10^{8}$\,G) and $\rm R_s$ is the NS radius; and the spatial extent 
$\rm x_0/R_{lc}$\footnote{
$\rm R_{lc}=c\,/\Omega=$96\,km
is the radius of the light cylinder of the pulsar in J1431} 
of the emitting region for particles injected in a
given injection point $\rm x_{in}$.
The parameter $\rm x_0/R_{lc}$ plays the role of a weighting function representing the
reduction of the number of emitting particles directed
towards the observer at a distance from their injection point.
The model solves the equation of motion of the particles to obtain their trajectories, 
and computes the emission of particles along each point. The latter is dependent on
the local kinetic properties, like Lorentz factor, the pitch angle and position, 
and the local properties of the magnetic field \citep[see details in][]{Iniguez22}. 

\noindent Fig.\,\ref{fig:sc-fit} (left panel) shows the results  of the model with 
$\rm log\,E_{||}=9.31 \pm 0.04\,V\,m^{-1}$, $\rm log\,(x_0/R_{lc})=-2.34 \pm 0.04$ and 
$b$= 1.24$\pm$0.23 ($\rm \chi^2_{\nu}$/d.o.f.=2.29/23).
The fit is not perfect, but reasonably close to the data.
The X-ray emission is the worse described by the model, 
and, related to that, the magnetic gradient is low in comparison with other 
fitted pulsars \citep{Iniguez22,Iniguez24}. This can come as a result of two related effects.
First, the X-ray spectrum is steeply rising at soft energies, more than usually found for 
other pulsars, and as a result the peak in X-rays is comparable to the peak in gamma-rays,
so that the model is trying to keep up with a large synchrotron emission.
Second, this may also come as an effect of considering a more extended region. 
Farther from the pulsar, the magnetic field still needs to be high --decay more slowly-- 
in order for synchrotron emission to proceed. In addition, 
the possible NS thermal contamination would mostly affect at these low energies. 

Given such possibility, the SED has also been fit excluding 
the soft X-ray portion of the spectrum, and hence below 2\,keV.  Fig.\,\ref{fig:sc-fit} (right panel)
shows the results with $\rm log\,E_{||}=9.56 \pm 0.04\,V\,m^{-1}$, 
$\rm log\,(x_0/R_{lc})=-2.59 \pm 0.04$ and
$b$=0.99$\pm$0.23 ($\rm \chi^2_{\nu}/d.o.f.$=0.72/7), which describe much better the
observed SED. The higher value of the parallel electric field and the lower $b$ are 
the consequence of the steep X-ray slope. Since there are not many MSPs with both X-ray and gamma-ray data 
\citep[see][]{Iniguez22}, it is not possible to state whether this steep spectrum is a common or unusual
property of MSPs. 

In summary, a magnetospheric origin for the X-ray to gamma-ray emission seems to be plausible.

\section{Conclusions}
\label{sec:conclusions}

We have analysed for the first time the X-ray and optical emission of the energetic
MSP pulsar binary J1431 to assess the system parameters and characterize its multi-band
emission. Here we summarise the main results:

\begin{itemize}
\item
 Our first deep X-ray observation of J1431 has detected a dim X-ray
source  without signficant  variability at its 10.8\,h orbital period. 
 This contrasts with the majority
of spiders displaying substantial X-ray orbital variations, which are often found double-peaked
due to Doppler boosting of the IBS synchrotron emission. 

\item
The X-ray spectrum is featureless and consistent with non-thermal emission
with a power law photon index $\Gamma$= 1.6$^{+0.3}_{-0.2}$ and negligible absorption 
($\rm N_h \leq 8\times10^{20}\,cm^{-2}$). While not statistically
significant, the presence of a thermal component ($\rm kT_{BB} =0.15\pm0.04$\,keV) may hint
at the contribution of the heated polar cap ($\rm R_{BB}\sim$ 180-660\,m) in the soft X-rays
together with a harder ($\Gamma=1.3\pm$0.4) power law component dominating at higher energies. 

\item 
Given that the X-ray timing and spectral analyses do not strongly favour
a  dominant contribution from an IBS, we inspected whether a magnetospheric origin from the MSP can describe
the X-ray and gamma-ray SED and found a reasonable match. 
This suggests that an IBS in this system, if present, has a negligible contribution.

\item

 The companion star is found to be an early F-type star with a mass of 0.20$\pm0.04\rm M_{\odot}$
which  underfils its Roche lobe ($f_{\rm RL}=73\pm4\%$), confirming the redback nature of J1431,
 although hotter than the majority of redbacks.
The multi-colour optical light curves  display typical 
ellipsoidal modulation at the long orbital period but are also affected by irradiation. 
Among the several models attempted to fit the multi-colour light curves and radial velocity
curve,  similar parameters are found with a model that includes direct heating from the 
pulsar and a cold spot with respect to the base stellar temperature ($\Delta T \sim -500$\,K),  
as well as with models that 
include heat redistribution after irradiation encompassing convection and also 
both diffusion and convection. The resulting
temperatures of the dayside and nightside faces of the companion are found to be $\sim$7500\,K
and 7400\,K, respectively, indicating a mild irradiation temperature $\rm T_{irr}\sim$3900\,K.
The lack of a strong irradiation is likely due to the high luminosity of the star which
is about 6.5 times larger than that expected from heating at the orbital distance 
from the pulsar. 

\item

 The binary inclination is found to be $i= 59\pm6^{\circ}$, explaining
the lack of orbital modulation in the X-rays and gamma-ray regimes. Furthermore, a 
distance of 3.1$\pm$0.3\,kpc is found, which locates J1431 at larger distance than
previously estimated.

\item

 Despite the inability to derive the best irradiation model, which limits
the precision to which the NS mass is constrained, a massive ($\rm 1.8-2.2\,M_{\odot}$)
NS in J1431 is favoured. 
\end{itemize}

\begin{acknowledgements}
This work is based on observations obtained with \XMM , am ESA science mission
with instruments and contributions directly funded by ESA Member States and
NASA and on observations collected at the 
European Southern Observatory under ESO programme 0103.D-0241.
This work has made use of data from SkyMapper, a facility funded through 
ARC LIEF grant LE130100104 from the Australian Research Council, 
awarded to the University of Sydney, the Australian National University, 
Swinburne University of Technology, the University of Queensland, the 
University of Western Australia, the University of Melbourne, Curtin 
University of Technology, Monash University and the Australian 
Astronomical Observatory. SkyMapper is owned and operated by The 
Australian National University's Research School of Astronomy and 
Astrophysics. 
This work has also  made use of data from the European Space Agency (ESA) mission
{\it Gaia} (\url{https://www.cosmos.esa.int/gaia}), processed by the {\it Gaia}
Data Processing and Analysis Consortium (DPAC,
\url{https://www.cosmos.esa.int/web/gaia/dpac/consortium}). Funding for the DPAC
has been provided by national institutions, in particular the institutions
participating in the {\it Gaia} Multilateral Agreement.
The \XMM data are publicly available at the European Space Agency (ESA) archive
http://nxsa.esac.esa.int/nxsa-web. The \UCAM data can be obtained by contacting
the corresponding author or the \UCAM team (V.S.\,Dhillon).  

\noindent DdM, AP, FCZ, AMZ acknowledge financial support from the Italian Institute for Astrophysics 
(INAF) Research Grant 2022:"FANS". AP is supported by the Italian Ministry
of University and Research (MUR), PRIN\,2020 (prot.2020BRP57Z) "Gravitational 
and Electromagnetic-wave Sources in the Universe with current and next generation
detectors (GEMS).  
FCZ is supported by a Ram\'on y Cajal fellowship (grant agreement RYC2021-03088-I).
VSD and \UCAM operations are funded by the Science and Technology Facilities Council (grant ST/Z000033/1).
RPB acknowledges support from the European Research Council (ERC) 
under the European Union's Horizon 2020 research and innovation program
(grant agreement N.715051; Spiders).
We also acknowledge the support of the PHAROS COST Action (CA\,16214).

\noindent We acknowledge useful comments from the anonymous referee.

\end{acknowledgements}
 

\bibliographystyle{aa}
\bibliography{biblio}

\FloatBarrier

\begin{appendix}


\section{Corner plots for the three best fit models with post-IGD}


\begin{figure*}
	\includegraphics[width=\textwidth]{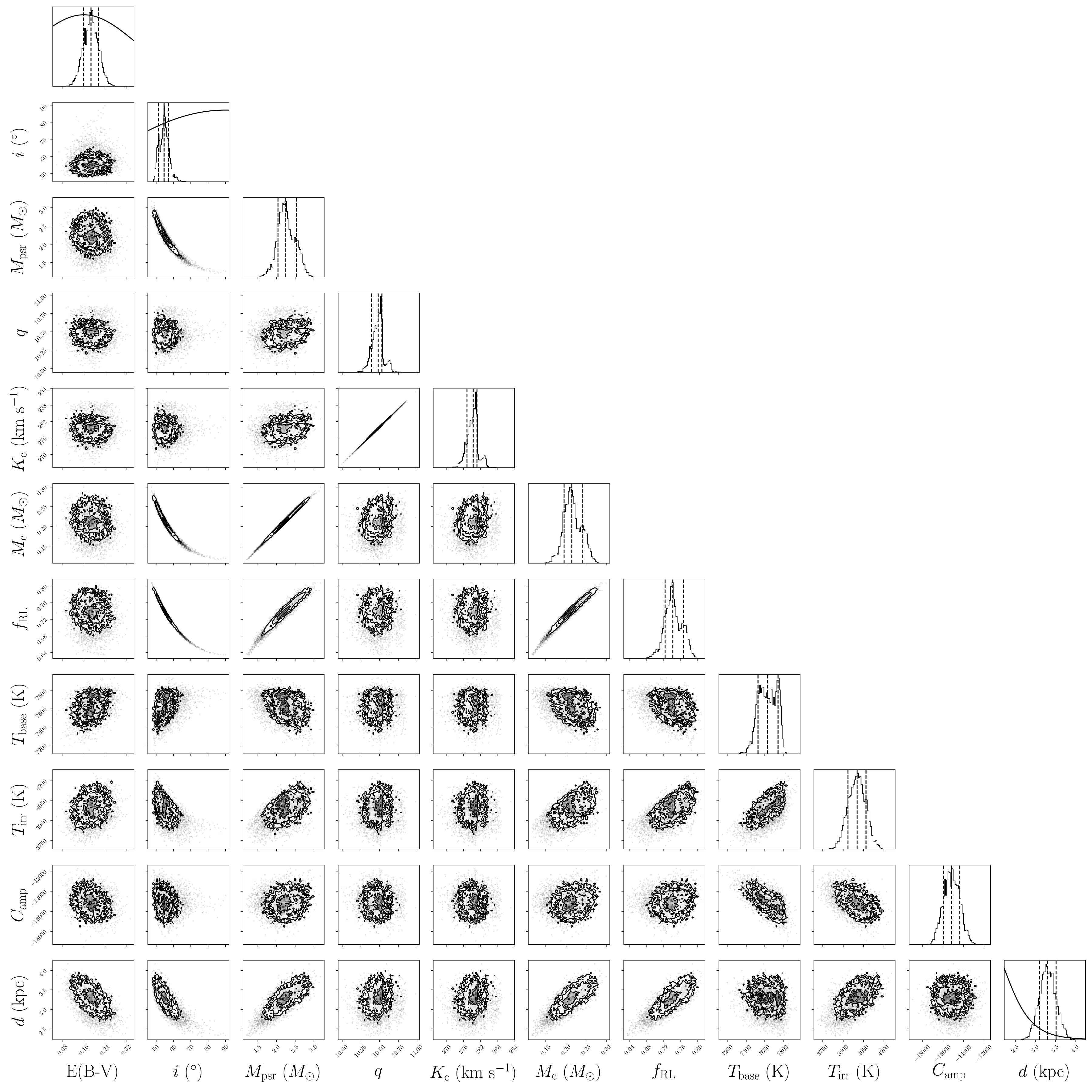}
    \caption{Corner plot displaying the fitted and derived parameters for the
convection (C) model with post-irradiation gravity darkening (post-IGD). 
Contours are the 68$\%$, 95$\%$ and 
99.7$\%$ confidence levels. The plots along the diagonal are the posterior 
distributions of the parameters and the solid lines overlaid are the prior 
distrbutions. The vertical dashed lines represent the 0.025, 0.5 and 0.975 quantiles.}
    \label{fig:cornerc}
\end{figure*}

\begin{figure*}
	\includegraphics[width=\textwidth]{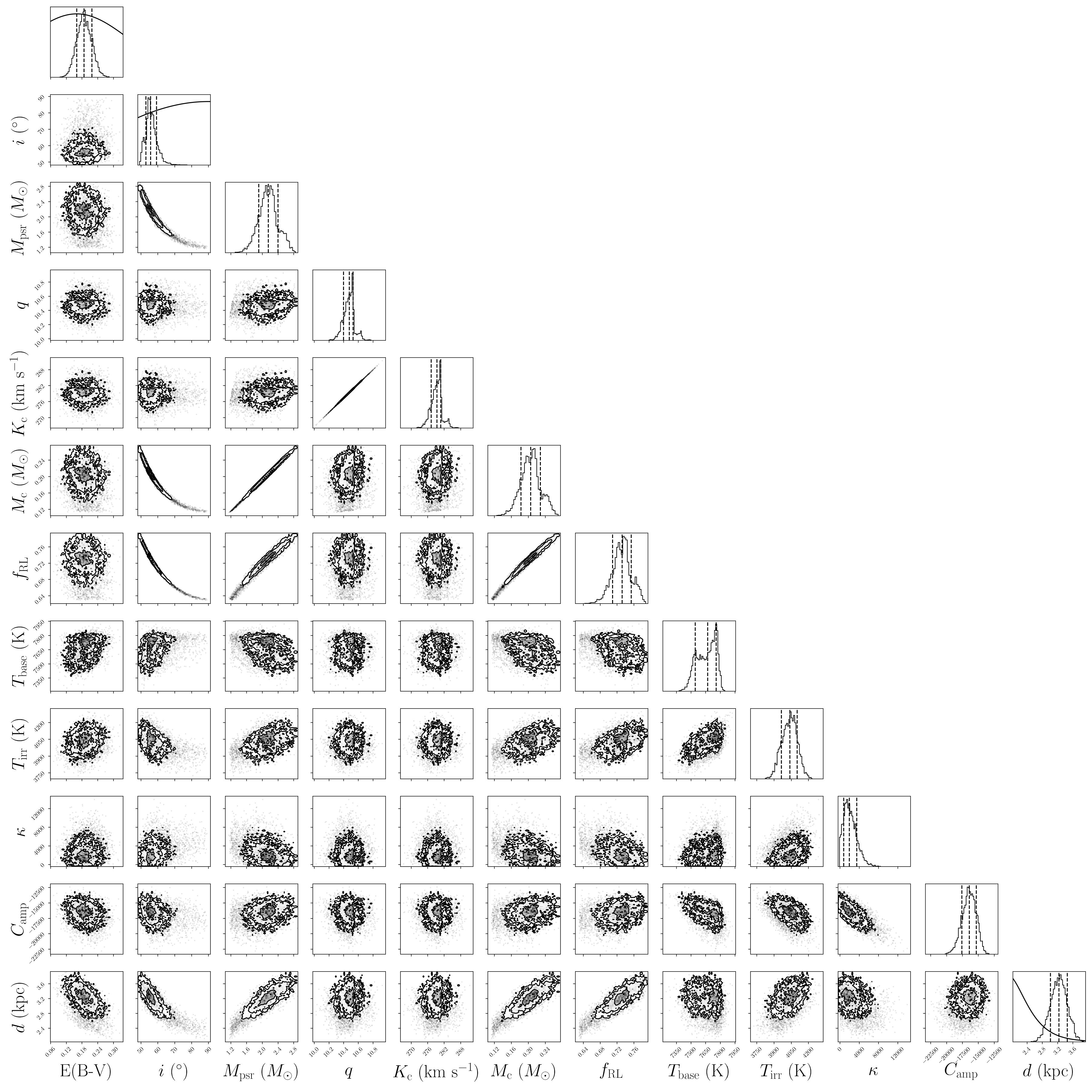}
    \caption{Corner plot displaying the fitted and derived parameters for the
diffuse plus convection (D+C) model with post-irradiation gravity darkening 
(post-IGD).  Contours are the 68$\%$, 95$\%$ and 
99.7$\%$ confidence levels. The plots along the diagonal are the posterior 
distributions of the parameters and the solid lines overlaid are the prior 
distrbutions. The vertical dashed lines represent the 0.025, 0.5 and 0.975 quantiles.}
    \label{fig:cornerdc}
\end{figure*}

\begin{figure*}
	\includegraphics[width=\textwidth]{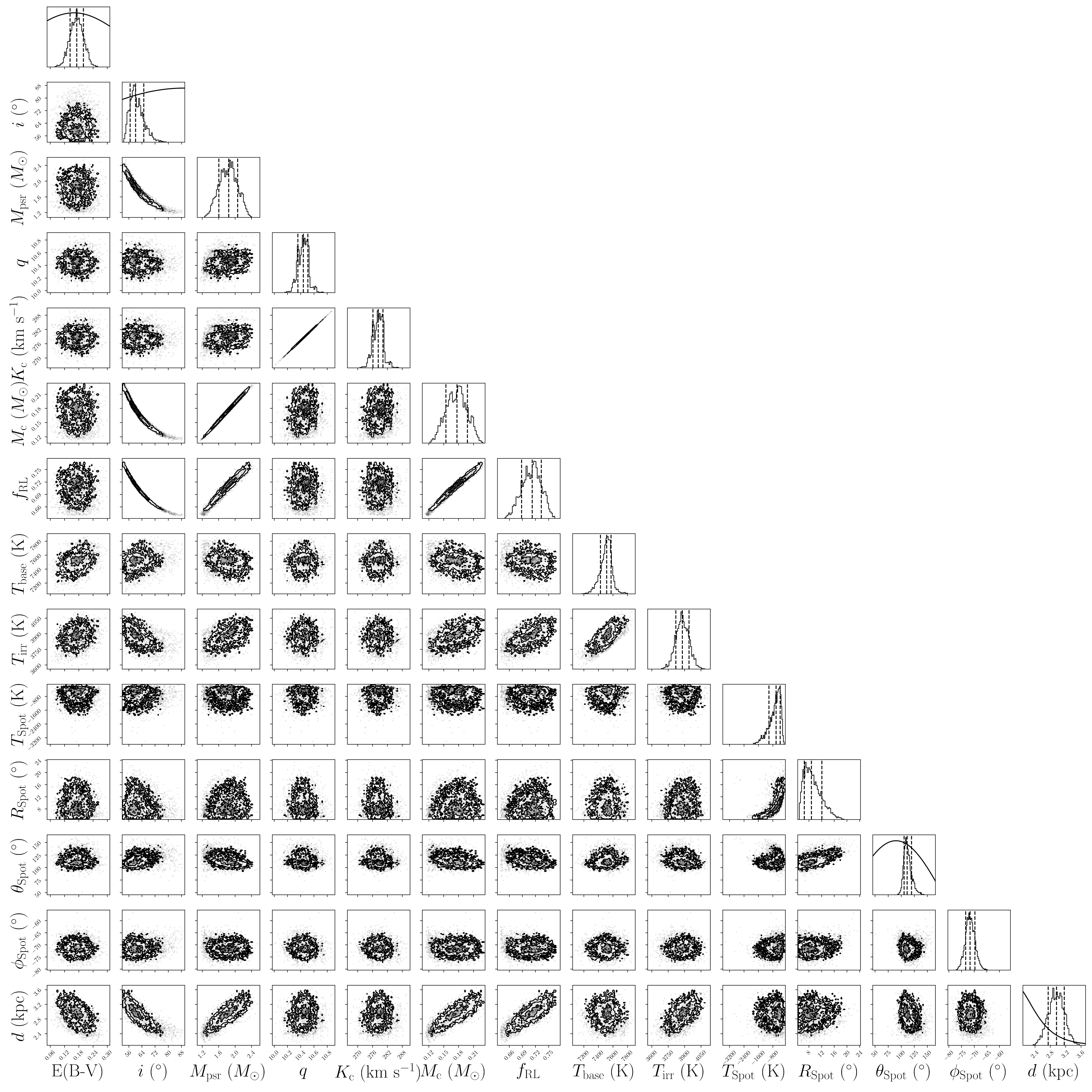}
    \caption{Corner plot displaying the fitted and derived parameters for 
the direct heating (DH) model with a cold spot and with post-irradiation gravity darkening
(post-IGD). Contours are the 68$\%$, 95$\%$ and 
99.7$\%$ confidence levels. The plots along the diagonal are the posterior 
distributions of the parameters and the solid lines overlaid are the prior 
distrbutions. The vertical dashed lines represent the 0.025, 0.5 and 0.975 quantiles.}
    \label{fig:cornerspot}
\end{figure*}

\end{appendix}

\end{document}